\documentstyle[preprint,aps,eqsecnum]{revtex}
\tightenlines
\begin{document}
\draft
\title{{\bf The $\pi$NNN--NNN problem. Connectedness, transition 
amplitudes and quasi--particle approximation}}
\author{G. Cattapan and L. Canton}
\address{Istituto Nazionale di Fisica Nucleare e 
Dipartimento di Fisica dell' Universit\'a\\
via Marzolo 8, Padova I-35131, Italia}

\maketitle

\begin{abstract}
      In this paper we review the present status of the $\pi$NNN--NNN
problem. In particular, we re--consider the chain--labelled
approach recently proposed by us, and identify a class of 
graphs, previously overlooked, which prevents the kernel of the 
corresponding $\pi$NNN--NNN equations from being connected. We propose 
some approximate schemes, yielding connected--kernel equations. 
A generalization of the residue method allows to relate the 
transition amplitudes for the coupled $\pi$NNN--NNN system to the 
chain--labelled formalism. The quasi--particle approach
is extended to the present situation, where emission/absorption
of particles is allowed. The open problems for the $\pi$NNN--NNN 
system in the light of the present and of previous approaches
are finally discussed. 
\end{abstract}

\pacs{PACS numbers: 21.45.+v, 11.80.Jy, 13.75.Gx, and 25.80.-e \\
\hfill Preprint DFPD 97/TH/17}


\narrowtext

\section{Introduction}
\label{intro}

      In two previous papers \cite{cc1,cc2} we have 
outlined a new approach to the $\pi$NNN--NNN problem, which can be regarded 
as the natural extension of the Grassberger--Sandhas (GS) n--body method
\cite{grass67,san80} to situations in which the number of particles is 
no longer conserved. As is well--known, the GS formalism represents 
a basic achievement in modern n--body scattering theory. It casts in 
a physically transparent form the connected--kernel Yakubovski\u{\i} 
(Y) scheme \cite{yak67}, which enjoys the distinctive feature of 
being completely free from non--physical solutions. Indeed, not only
it guarantees that the Fredholm alternative holds, so that non--physical 
solutions do not contaminate the physical one in the scattering region, 
but it satisfies also a constrained Fredholm alternative \cite{ak91}; 
the non--trivial solutions of the associated homogeneous equations are 
in one--to--one correspondence with the bound states of the total system. 
Most of the connected--kernel n--body equations proposed
in the literature satisfy the former condition, but fail in guaranteeing 
the latter. 

      The extension of the GS formalism considers scattering and
pion production/absorption processes on equal footing, thus
providing a coupled treatment of all the relevant processes in the
same dynamically consistent framework. The $\pi$ production introduces
radical changes in the structure of the dynamical equations and
gives rise to challenges and problems which were not possible in
standard n--body theories.

      On a fundamental level, there is the problem with 
nucleon renormalization, which is unavoidable in theories
where the Fock space is truncated to states with at most one
pion. This problem is well acknowledged in the literature
\cite{sauer85,bla92,kb93,bk94,kbrel94,pa95,pa96}, and herein 
is not discussed.

      The first problem considered in this paper concerns the 
connectedness of the $\pi$NNN integral equations, and we arrive
at the conclusion that, contrary to our previous belief \cite{cc2},
these equations are not connected. In this paper we identify a
class of graphs, previously overlooked, which prevents the kernel
from being connected. We show that the problem arises when the
$3+1$ (($\pi$NN)N) and $2+2$ ((NN)($\pi$N)) partitions are treated
on equal footing, and that connectedness can be restored with two
iterations of the kernel if the coupling between the $\pi$NNN and
NNN sectors is switched off in the $2+2$ partitions, whereas it is 
treated exactly in the $\pi$NN sub--systems. As a matter of fact,
connectedness is guaranteed also when production/absorption
processes are allowed in $2+2$ partitions, provided that the
exact sub--system amplitudes are projected in the sole $\pi$NNN
sector.

      The second problem we analyze in detail is the problem with
the identification of certain physical transition amplitudes which
refer to the nucleon--deuteron channel. The ambiguities in the
definition of these amplitudes are originated by the delicate
interplay between the $2+2$ and $3+1$ partitions, and are closely
related with the connectedness problem. Indeed, with the same
approximation scheme we restore the connectedness in our equations
and solve these amplitude ambiguities.

      The extension of the GS approach to the $\pi$NNN system 
is briefly illustrated in Sect. \ref{revrev}. Starting from the 
transition amplitudes for $4\rightarrow4$ processes, one first extracts 
operators referring to three--cluster$\rightarrow$ three--cluster 
transitions, which satisfy dynamical equations formally identical to 
the Afnan--Blankleider (AB) equations for the $\pi$NN--NN system 
\cite{afnan80,avi83,ccs}. By resorting to the powerful GS matrix technique, 
the two--cluster partitions are introduced into the theory, and one gets 
at the end dynamical equations quite similar in structure to the GS 
four--body equations. The presence of emission/absorption processes, however, 
implies some noticeable differences, when these equations are analyzed 
in detail. The chain--of--partition labelling, characterizing the YGS 
approach, is now more complex, since one has to regard nucleon pairs 
both as three--cluster partitions in the $\pi$NNN sector, and as
two--cluster partitions in the NNN space. As a consequence, one deals with
a set of 24 coupled equations, in place of the 18 characterizing
the standard YGS scheme. 

      In Sect. \ref{res} we establish the relation between the employed 
operators, labelled by chains of partitions, and the physical amplitudes 
connecting the various partitions of the system. As is well-known in 
non--relativistic scattering theory, the physical transition amplitudes 
can be identified by exhibiting the momentum--space singular terms of 
the total Green function, in correspondence to the bound states of the 
various sub--systems \cite{ak91,fad65,ro65}. This procedure, which has its 
relativistic counterpart in reduction formulae of Quantum Field 
Theory \cite{itz85}, is often referred to as the residue method
\cite{ak91,afnan80}. Alternatively, one can work in the language of 
transition operators. For the standard n--body problem, by combining 
the residue rule with the GS matrix technique, it has been shown 
\cite{cc2} that the two--cluster$\rightarrow$two-cluster transition 
amplitudes can be related to the solution of the n--body YGS equations, 
starting from the transition amplitudes for n$\rightarrow$n processes.
Here, as anticipated in \cite{cc1}, we generalize this procedure
to the $\pi$NNN--NNN case, where the coupling between spaces
with different numbers of particles has to be taken into account.
Starting from the relations expressing the $4\rightarrow4$ transition 
amplitudes in terms of operators referring to less and less clusterized 
partitions, we systematically identify the poles associated to the 
bound--states or resonances in the various two-- and three--body 
sub--systems. As the outcome of this procedure, we express both the 
rearrangement and the meson absorption/emission transition amplitudes 
in terms of the solution of our chain--labelled equations.
Finally, in the same Section, we discuss also the ambiguities in the
identification of the physical amplitudes, and the associated role
played by the $3+1$ and $2+2$ partitions in this problem.

      In Sect. \ref{qps} we consider the Quasi--Particle--Approximation 
(QPA). In the four--body case, this amounts to a two--step procedure 
\cite{grass67,san80,ak91}. One first replaces the two--body t--matrices 
with finite--rank operators, thereby re--writing the YGS equations as 
effective Faddeev equations for two elementary particles and a composite 
object; as a second step, one approximates again the three--body and 
$2+2$ sub--system amplitudes by finite--rank operators. One thus obtains 
multichannel Lippmann--Schwinger--type (LS) equations in one vector 
variable, coupling the elastic/rearrangement transition amplitudes. The
physical transition amplitudes for break--up processes can be 
evaluated starting from the solution of the effective LS equations
through simple quadratures. We show here that the QPA can be extended to 
the present situation. Here also the two--fragment$\rightarrow$two--fragment
transition amplitudes satisfy LS equations, all the break--up and 
emission/absorption amplitudes being expressible in terms of them by 
quadratures. The QPA approach is particularly attractive because it
shows the theory through diagrams which can be easily interpreted.
It is then possible to view the diagrams which are at the origin of
the disconnectedness problem. These disconnected graphs are 
self--energy--type contributions to the transition from a ($\pi$N)(NN)
to a N($\pi$NN) configuration with intermediate zero--pion state.
If this term is disregarded, one gets connected--kernel 
equations, much as in the standard four--body problem. Needless to say, 
the same is true if emission/absorption processes are switched off 
in $2+2$ sub--systems, or allowed only in the sub--system amplitudes
referring to the four--body sector.

      Given the present situation, we review in Sect. \ref{finis} 
the approaches available in literature \cite{avi83,ueda}. In 
\cite{avi83} one applies the quasi--particle approximation 
to the basic AB equations, and removes the disconnected pieces occurring 
in the standard four--body problem first, by resorting to the usual
GS method. All disconnected terms due to pion emission/absorption 
are then treated together, by use of a two--potential formula.
Disconnected contributions are formally regarded as an auxiliary problem,
whose solution gives the input for connected--kernel equations
yielding the physical transition operators. As a consequence,
the actual transition amplitudes can be evaluated only solving
nested sets of integral equations. Finally, these rigorous FY--type 
approaches are compared to the effective, coupled--channel formalism of 
Ref. \cite{ueda}. It turns out that this phenomenological method is at 
variance with present few--body scattering theory, since it assumes 
{\it ad hoc} couplings between configurations of the $\pi$NNN and NNN 
systems, which are excluded in more microscopic FY--type formulations.

\section{The dynamical equations}
\label{revrev}

      As in Refs. \cite{cc1,cc2}, we start from the unclusterized transition 
operators which are allowed in the NNN and $\pi$NNN sectors of the 
theory. Let $T_{(0|0)}$ be the operator associated to the transition from
a three--nucleon state $|\chi_{0}>$ to a three--nucleon state 
$|\chi_{0}'>$, so that the corresponding transition amplitude is given
by $<\chi_{0}'|T_{(0|0)}|\chi_{0}>$. Similarly, $T_{(1|1)}$ describes a
transition from an initial $\pi$NNN state $|\chi_{1}>$ to a final one
$|\chi_{1}'>$. The two sectors communicate through the absorption and
production operators $T_{(0|1)}$ and $T_{(1|0)}$, respectively, the
associated transition amplitudes being $<\chi_{1}'|T_{(1|0)}|\chi_{0}>$,
and $<\chi_{0}'|T_{(0|1)}|\chi_{1}>$. The AB transition
operators can be introduced through the relations \cite{cc1,cc2}
\begin{mathletters}
\label{aball}
\begin{eqnarray}
T_{(0|0)} &=& U \label{aba}, \\
T_{(1|0)} &=& \sum_a t_a G_0 U_a \label{abb}, \\
T_{(0|1)} &=& \sum_b U_b^\dagger G_0 t_b 
\label{abc}, \\
T_{(1|1)} &=& \sum_a t_a + \sum_{a,b}t_aG_0U_{ab}G_0t_b \label{abd} .
\end{eqnarray}
\end{mathletters}
      Here, $G_0$ represents the free $\pi$NNN propagator. As usual in 
few--body scattering
theory, indices such as $a$, $b$, and $c$ denote generic three--cluster 
partitions of the four--body system, namely, interacting pairs in presence 
of two spectator particles. If necessary, to distinguish between NN and 
$\pi$N pairs, we shall denote the latter by $i$, $j$ or $k$, so that
$i$ represents the pair $\pi{\rm N}_i$ with nucleons ${\rm N}_j$ and
${\rm N}_k$ as spectators ($i,j,k$ a cyclic permutation of $1,2,3$). 
Finally, the operators $t_a$ are the NN or $\pi$N t--matrices. 
To be consistent with the explicit allowance of the $\pi$NN vertices,
only the non--polar part has to be retained in the $\pi$--nucleon
t--matrices $t_i$ in the $P_{11}$ channel\cite{afnan80}. For the
sake of simplicity, we have omitted the dependence upon the energy
variable $z$ in the resolvent and transition operators. It will be
exhibited only when necessary. For the same reason, the outgoing
boundary conditions assumed on--shell for the operators are not
explicitly indicated. Note that the operators which are obtained
through Hermitean conjugation are associated to ingoing boundary
conditions, so that one has $U^\dagger_b \equiv U_b(E-i0)^\dagger$,
with $E$ the total energy.

      The physical meaning of the operators introduced through
eqs.~(\ref{aba}--\ref{abd}) can be ascertained by studying the
behaviour of the $4\rightarrow 4$ amplitudes in momentum--space,
near the poles corresponding to the two--body bound states or
resonances. As is well--known, the dominant part of $t_a$ at the 
bound--state (or resonance) energy $z=E_a$ can be written \cite{note1}
\begin{equation}
t_a \simeq |a> {1 \over z - E_a} <a| ,
\label{pole}
\end{equation}
where $|a>$ is the form factor for the correlated pair $a$,
\begin{equation}
|a> = V_a |\phi_a> .
\end{equation}
Here, $V_a$ represents the two--body interaction in pair $a$, and
$|\phi_a>$ satisfies the homogeneous equation 
\begin{equation}
G_0(E_a)V_a|\phi_a> = |\phi_a> .
\label{chaneq}
\end{equation}

      For $z \sim E_a$ the production amplitude can then be
written, because of Eq.~(\ref{abb})
\begin{equation}
<\chi_{1}'|T_{(1|0)}(z)|\chi_{0}> \simeq <\chi_{1}'|V_a|\phi_a>{1\over z-E_a}
<\phi_a|U_a|\chi_{0}> .
\label{pol1}
\end{equation}
The residue of the amplitude at this simple pole provides (apart from
the form factor \break $<\chi_{1}'|V_a|\phi_a>$) the transition amplitude
$<\phi_a|U_a|\chi_{0}>$ referring to three--cluster$\rightarrow$three--cluster
transitions from the NNN space to the $\pi$NNN sector. Similarly, one can 
establish that $U^\dagger_b$ and $U_{ab}$ are absorption and reaction 
operators, respectively, for transitions between three--cluster 
configurations of the $\pi$NNN--NNN system.

      The dynamical equations for the AB operators have been derived
in Ref. \cite{afnan80} for the $\pi$NN system, by resorting to 
Taylor's diagrammatic method \cite{tay,bro}, and have been extended
to the $\pi$NNN case in refs.\cite{avi83,ccs}. For the $\pi$NNN case 
these equations do not have a connected kernel,
so that they may have unphysical solutions in addition to the correct
(physical) one. In our previous papers \cite{cc1,cc2} we have shown that 
Faddeev--Yakubovski\u{\i}--type (FY) equations, and explicit allowance for 
two--cluster partitions can be obtained through a non--trivial generalization 
of the GS approach to the four--body problem
\cite{grass67,san80}. To this end one writes the four--body AB
equations in the matrix form
\begin{equation}
{{\sf T}^{(3)}} = {{\sf V}^{(3)}} + 
{{\sf V}^{(3)}}{\sf G}_0^{(3)}{{\sf T}^{(3)}} ,
\label{lsab}
\end{equation}
where ${\sf G}_0^{(3)}$, ${{\sf V}^{(3)}}$ and ${{\sf T}^{(3)}}$ are
matrices in the three--cluster--partition indices defined according to
\begin{eqnarray}
{\sf G}_{0}^{(3)}
\equiv & 
\left|       
\begin{array}{cc}
{G_{0}^{(3)}}_{\!\!(a|b)} & {G_{0}^{(3)}}_{\!\!(a|0)} \cr 
{G_{0}^{(3)}}_{\!\!(0|b)} & {G_{0}^{(3)}}_{\!\!(0|0)} \cr
\end{array} \right| 
&= 
\left| \begin{array}{ccc}
G_0t_aG_0\delta_{ab} &\ & 0 \\
0 &\ & g_0 \\
\end{array} \right|,
\label{prop}\\
{{\sf V}}^{(3)}
\equiv 
& \left| 
\begin{array}{cc}
V^{(3)}_{(a|b)} & V^{(3)}_{(a|0)} \cr 
V^{(3)}_{(0|b)} & V^{(3)}_{(0|0)} \cr
\end{array} \right| 
&= 
\left| \begin{array}{ccc}
G_0^{-1}\bar\delta_{ab} &\ & F_a\\
F^\dagger_b &\ & {\cal V}\\
\end{array} \right| ,
\label{pott}\\
{{\sf T}}^{(3)}
\equiv 
& \left| 
\begin{array}{cc}
T^{(3)}_{(a|b)} & T^{(3)}_{(a|0)} \cr 
T^{(3)}_{(0|b)} & T^{(3)}_{(0|0)} \cr
\end{array} \right| 
&= 
\left| \begin{array}{ccc}
U_{ab} &\ & U_a\\
U^\dagger_b &\ & U \\
\end{array} \right| ,
\label{titre}
\end{eqnarray}
respectively. The notation exhibits the fact that the diagonal 
blocks of these matrices refer to the $\pi$NNN and NNN spaces, whereas the 
off--diagonal blocks contain the operators connecting the two sectors.
Here, $g_0$ is the free three--nucleon propagator, and ${\cal V}$ represents 
the total interaction in the NNN sector with the one--pion--exchange 
contributions explicitly included; $F_a$ ($F_b^\dagger$) is the sum of 
the elementary production (absorption) vertices external to the pair $a$ 
(see Eq. (1) in Ref. \cite{cc1}), and $\bar\delta_{ab} \equiv 1 - \delta_{ab}$.

      As in the standard GS approach, a crucial step in obtaining
equations with a FY coupling scheme is represented by a sum rule, by which
the matrix interaction ${{\sf V}^{(3)}}$ is written as the sum of
contributions ${\sf v}_{a'}^{(3)}$ referring to the two--cluster 
partitions $a'$ of the total system. This can be accomplished through
a more refined classification of the two--cluster partitions, with
respect to the usual four--body theory. If $a'$ is the partition
$\pi({\rm NNN})$, with the pion a mere spectator (Type--III partition),
one essentially has the GS form \cite{grass67,san80}
\begin{mathletters}
\begin{equation}
{\sf v}_{a'}^{(3)}
\equiv \left| 
\begin{array}{cc}
{v_{a'}^{(3)}}_{\!\!(a|b)} & {v_{a'}^{(3)}}_{\!\!(a|0)} \cr
{v_{a'}^{(3)}}_{\!\!(0|b)} & {v_{a'}^{(3)}}_{\!\!(0|0)} \cr
\end{array} \right| = 
\left| \begin{array}{cc}
{G_0}^{-1}\bar\delta_{ab}\delta_{a,b\subset a'} & 0 \\
                      0                         & 0 \\
\end{array} \right|,
\label{vags}
\end{equation}
with $\delta_{a,b\subset a'}$ equal to one if both $a$ and $b$ are 
obtained by breaking a cluster in $a'$, and zero otherwise.
If, on the other hand, $a'$ contains an interacting $\pi$NN system 
plus a spectator nucleon (Type--I partition) or two pairs $\pi$N and 
NN with no mutual interaction (Type--II partition) one has 
\begin{equation}
{\sf v}_{a'}^{(3)}
\equiv \left| 
\begin{array}{cc}
{v_{a'}^{(3)}}_{\!\!(a|b)} & {v_{a'}^{(3)}}_{\!\!(a|0)} \cr 
{v_{a'}^{(3)}}_{\!\!(0|b)} & {v_{a'}^{(3)}}_{\!\!(0|0)} \cr
\end{array} \right| = 
\left| \begin{array}{cc}
{G_0}^{-1}\bar\delta_{ab}\delta_{a,b\subset a'} & (f_{a'})_{a}  \\
                 (f^\dagger_{a'})_{b}           & {\cal V}_{a_1}\\
\end{array} \right| ,
\label{vab1}
\end{equation}
and 
\begin{equation}
{\sf v}_{a'}^{(3)}
\equiv \left| 
\begin{array}{cc}
{v_{a'}^{(3)}}_{\!\!(a|b)} & {v_{a'}^{(3)}}_{\!\!(a|0)} \cr 
{v_{a'}^{(3)}}_{\!\!(0|b)} & {v_{a'}^{(3)}}_{\!\!(0|0)} \cr
\end{array} \right| = \left| \begin{array}{cc}
{G_0}^{-1}\bar\delta_{ab}\delta_{a,b\subset a'} & (f_{a'})_{a}\\
           (f^\dagger_{a'})_{b}               &        0    \\
\end{array} \right| ,
\label{vab2}
\end{equation}
\end{mathletters}
respectively. Here, $(f_{a'})_{a}$ and $(f^\dagger_{a'})_{b}$ are emission and
absorption vertices internal to $a'$; they can be written in terms of 
the elementary production and absorption vertices for the $i$--th 
nucleon $f(i)$ and $f(i)^\dagger$, respectively, as follows
\begin{equation}
(f_{a'})_{a}={\sum_{i=1}^{3}}\bar\delta_{ia}\delta_{i,a\subset a'}f(i)
\ \ \ \ \ \ 
(f^\dagger_{a'})_{b}={\sum_{i=1}^{3}}\bar\delta_{ib}\delta_{i,b\subset a'}
f(i)^\dagger  .
\label{internalvert}
\end{equation}
The operator ${\cal V}_{a_1}$ represents the interaction internal to
the NN pair $a_1$ in the considered partition $a'$, with the 
one--pion--exchange tail included \cite{cc1,cc2}. We observe that $a_1$ 
represents at the same time a two--cluster partition in the three--nucleon 
sector and a three--cluster partition in the four--body space. It is uniquely
defined for each I-- or II--type $a'$. It is worth to note that, had we 
defined ${\sf v}_{a'}^{(3)}$ in the same way for Type--I and Type--II 
partitions, we would have counted the NN potentials ${\cal V}_{a_1}$
twice when summing ${\sf v}_{a'}^{(3)}$ over $a'$, to get the total
``interaction" ${\sf V}^{(3)}$.

      The operators ${\sf v}_{a'}^{(3)}$ represent the driving terms
of the equations for the sub--system dynamics. These equations can be 
written in the compact Lippmann--Schwinger form \cite{cc1,cc2}
\begin{equation}
{\sf t}^{(3)}_{a'}={\sf v}^{(3)}_{a'}+{\sf v}^{(3)}_{a'}{\sf G}_0^{(3)} 
{\sf t}^{(3)}_{a'} ,  
\label{subLS}
\end{equation}
where ${\sf t}^{(3)}_{a'}$ are matrices in the three--cluster--partition
indices, whose definition depends again upon the type of two--cluster
partition $a'$. For a Type-I or Type-II $a'$ they are defined according
to
\begin{mathletters}
\begin{equation}
{\sf t}_{a'}^{(3)}
\equiv \left| 
\begin{array}{cc}
{t_{a'}^{(3)}}_{\!\!(a|b)} & {t_{a'}^{(3)}}_{\!\!(a|0)} \cr 
{t_{a'}^{(3)}}_{\!\!(0|b)} & {t_{a'}^{(3)}}_{\!\!(0|0)}  \cr
\end{array} \right| = \left| \begin{array}{cc}
(u_{a'})_{ab} & (u_{a'})_{a}\\
(u^\dagger_{a'})_{b}& u_{a'} \\
\end{array} \right|,
\end{equation}
where $(u_{a'})_{ab}$, $(u_{a'})_{a}$, $(u^\dagger_{a'})_{b}$ and
$u_{a'}$ are AB--type transition operators describing scattering
and absorption/emission processes within the sub--systems defined
by $a'$. If, on the other hand, one has the Type--III partition 
$a'=\pi ({\rm NNN})$ one has the usual Alt--Grassberger-Sandhas 
(AGS) operators for the three--nucleon sub--system, namely 
\begin{equation}
{\sf t}_{a'}^{(3)}
\equiv \left| 
\begin{array}{cc}
{t_{a'}^{(3)}}_{\!\!(a|b)} & {t_{a'}^{(3)}}_{\!\!(a|0)} \cr
{t_{a'}^{(3)}}_{\!\!(0|b)} & {t_{a'}^{(3)}}_{\!\!(0|0)} \cr
\end{array} \right| = 
\left| \begin{array}{cc}
{(u_{a'})}_{ab} & 0 \\
        0       & 0 \\
\end{array} \right| .
\label{tags}
\end{equation}
\end{mathletters}

      The FY--type equations for the full $\pi$NNN system can 
be derived from the four--body AB equations (\ref{lsab})
by resorting to the basic ansatz
\begin{equation}
{{\sf T}^{(3)}}=\sum_{a'} 
{{\sf t}^{(3)} _{a'}} +\sum_{a'b'} 
{{\sf t}^{(3)}_{a'}}
{\sf G}_0^{(3)} 
{{\sf U}^{(3)}_{a'b'}}
{\sf G}_0^{(3)} 
{{\sf t}^{(3)}_{b'}} .
\label{fundamental}
\end{equation}
Requiring that ${\sf T}^{(3)}$, as given by (\ref{fundamental}), 
satisfies Eqs.~(\ref{lsab}) one gets
\begin{equation}
{{\sf U}^{(3)}_{a'b'}} =\bar\delta_{a'b'}{{\sf G}_0^{(3)}}^{-1}+
\sum_{c'}\bar\delta_{a'c'}{\sf t}^{(3)}_{c'}
{{\sf G}_0^{(3)}} {{\sf U}^{(3)}_{c'b'}} .
\label{yaku}
\end{equation}
These equations, once explicitly written, couple operators labelled by 
chains of partitions. Differently from the standard four--body case, 
however, now the coupling between spaces with different numbers
of particles is allowed; as a consequence, one has standard chain
indices $(a'a)$ (with $a\subset a'$) in the $\pi$NNN sector, and
hybrid--chain indices $(a'a_1)$ (with $a_1\subset a'$) to account
for the three--nucleon space. The explicit form of Eqs.~(\ref{yaku})
has been given and discussed elsewhere \cite{cc2}. Here, we limit
ourselves to observe that they couple the following operators:
operators $U_{a'ab'b}$ associated to scattering in the $\pi$NNN
space; operators $U_{a'a_1b'b_1}$ for collision processes in the
NNN sector; production and absorption operators $U_{a'ab'b_1}$ 
and $U^\dagger_{a'a_1b'b}$, respectively, which connect the two spaces
to each other.

      In two previous papers \cite{cc1,cc2}, we have analyzed the
connectedness properties of Eqs.~(\ref{yaku}). For all the graphs
considered therein, we found that their kernel ${\sf K}$ is connected 
after three iterations. A closer inspection, however, reveals a further
class of graphs, whose presence prevents the present formalism
from achieving connectedness. These graphs are related to 
self--energy insertions in Type--II partitions, by which
a nucleon line is dressed in presence of an interacting NN pair
[see Fig.~\ref{fig1}(a)]. Because of these contributions, disconnected
graphs appear in any iteration of ${\sf K}$, a typical disconnected term 
in ${\sf K}^4$ being exhibited in Fig.~\ref{fig1}(b). As already observed 
in ref.~\cite{avi83}, in a properly mass--renormalized theory these 
contributions to the NN interactions would never arise, since the
spectator nucleon would have already acquired its physical mass. 
In this truncated formalism, however, the FY coupling scheme implies 
that $2+2$ partitions have to be treated on equal footing as the 
Type-I ones, and one is forced to introduce the above dangerous graphs.

     On the ground of the above considerations, one has to
conclude that the present approach, in spite of its striking
similarities with the GS method, which has been so successful
in the standard four--body case, is not able to solve the 
disconnectedness problem for the $\pi$NNN--NNN system. A natural
question is whether one can extract from (\ref{yaku}) sensible
approximations, leading to connected--kernel equations. A first
possibility is suggested by the very nature of the disconnected
terms, namely, one can switch off the coupling between the
$\pi$NNN and NNN sectors for the Type--II partitions, thereby
allowing only multiple rescattering in the two--body sub--systems,
much as in standard four--body theory. This implies that the 
sub--system dynamics is described by Eqs.~(\ref{subLS}),
(\ref{vags}) and (\ref{tags}) not only for Type--III but also for 
Type--II partitions. Since hybrid chains are now introduced only for 
Type--I partitions, in this approximation Eqs.~(\ref{yaku}) represent
a set of 21 coupled equations. Proceeding as in Ref. \cite{cc2}, one
can verify that their kernel is connected after two iterations.
This approximation, therefore, gives an embedding of the AB
treatment for the $\pi$NN--NN sub--systems, into a four--body
approach, where the remaining part of the problem is handled
through a conventional multiple--scattering treatment. 

      A less severe truncation of the theory is possible, in
which pion emission and absorption can be allowed in $2+2$
partitions. To see how this
can be achieved, we write Eqs.~(\ref{subLS}) for the
scattering and absorption operators $(u_{a'})_{ab}$ and 
$(u^\dagger_{a'})_b$, with $a'$ a Type-II partition $(\pi i)(jk)$,
\begin{mathletters}
\begin{eqnarray}
(u_{a'})_{ab} 
& = &
G_0^{-1}\bar\delta_{ab} + \sum_{c} \bar\delta_{ac} t_c G_0 (u_{a'})_{cb} 
+ (f_{a'})_a g_0 (u^\dagger_{a'})_b \label{peq} \\
(u^\dagger_{a'})_b 
& = &
(f^\dagger_{a'})_b + \sum_{c} (f^\dagger_{a'})_c G_0 t_c G_0 (u_{a'})_{cb} .
\label{seq}
\end{eqnarray}
\end{mathletters}
Substituting for $(u^\dagger_{a'})_b$ in the former of these
equations from the latter, one gets
\begin{eqnarray}
(u_{a'})_{ab} 
& = &
G_0^{-1}\bar\delta_{ab} + (f_{a'})_a g_0 (f^\dagger_{a'})_b 
+ \sum_{c} \bar\delta_{ac} t_c G_0 (u_{a'})_{cb} \nonumber \\ 
& + & 
(f_{a'})_a g_0 \sum_{c} (f^\dagger_{a'})_c G_0 t_c G_0 (u_{a'})_{cb} .
\label{solveq}
\end{eqnarray}
One now assumes again Eq.~(\ref{tags}) for Type--II partitions.
In this approximation, emission/absorption processes contribute
to all orders to the $2+2$ sub--amplitudes in the $\pi$NNN
sector, while their components for direct pion emission/absorption
are set to zero. As a consequence, the intertwining between $3+1$
and $2+2$ partitions with zero--pion intermediate states is
forbidden, and connectedness can be achieved again.

\section{The transition amplitudes}
\label{res}
 
      The physical transition amplitudes for the various collision
processes allowed in the considered state space (${\cal H}_{{\rm NNN}}
\oplus {\cal H}_{\pi{\rm NNN}}$) can be extracted from 
Eqs.~(\ref{aball}) and (\ref{fundamental}) through a suitable
generalization of the residue method. To exhibit the singularities 
of the relevant operators in correspondence to the possible bound 
(or resonant) states of the sub--systems, we consider the 
homogeneous eigenvalue problem associated to Eqs.~(\ref{subLS}), 
namely
\begin{equation}
|{\bf \Gamma}^{(3)}_{a'}(E_{a'})> = {\sf v}^{(3)}_{a'}(E_{a'})
{\sf G}_0^{(3)}(E_{a'})|{\bf \Gamma}^{(3)}_{a'}(E_{a'})> .
\label{homo1}
\end{equation}
For Type--I or Type--II partitions, $|{\bf \Gamma}^{(3)}_{a'}(E_{a'})>$ 
is a column vector in the space of the three--cluster partitions of both
sectors of the model state space. In particular, in the $\pi {\rm NNN}$ 
sector, $|{\bf \Gamma}^{(3)}_{a'}>$ has non--vanishing components 
$|\Gamma_{a'a}>$ in those three--cluster partitions obtained from
a sequential break--up of the two--cluster partition $a'$. For the 
Type--III partition, the column vector has non--vanishing 
components $|\Gamma_{a'a}>$ only in the $\pi$NNN sub--space. For the sake 
of simplicity, we have assumed that there is a unique eigenvalue for a 
given $a'$, the generalization to many eigenvalues $E_{a'r}$ implying 
only a more involved bookkeeping of indices. If the state--vectors
\begin{equation}
|{\bf \Phi}^{(3)}_{a'}(E_{a'})> \equiv {\sf G}_0^{(3)}(E_{a'})
|{\bf \Gamma}^{(3)}_{a'}(E_{a'})> 
\label{wfun}
\end{equation}
are introduced, Eq.~(\ref{homo1}) becomes
\begin{equation}
|{\bf \Phi}^{(3)}_{a'}(E_{a'})> = {\sf G}_0^{(3)}(E_{a'})
{\sf v}^{(3)}_{a'}(E_{a'})|{\bf \Phi}^{(3)}_{a'}(E_{a'})> .
\label{homo2}
\end{equation}

      To have a first insight into the physical meaning of these 
equations, it is instructive to write Eq.~(\ref{homo2}) explicitly.
For the Type--III partition, taking the Eqs.~(\ref{prop}) and
(\ref{vags}) into account, one gets
\begin{equation}
|\Phi_{a'a}> = G_0t_a\sum_{c(\subset a')}\bar\delta_{ac}
|\Phi_{a'c}> ,
\label{wfags}
\end{equation}
namely, the standard Faddeev equations for the wave--function components
$|\Phi_{a'a}>$ of the three--nucleon bound state in presence of the
spectator pion. As is well--known, the total wave function is given by
$\sum_{a}|\Phi_{a'a}>$.

      For the Type--I partitions one has from Eqs. (\ref{prop}) and
(\ref{vab1})
\begin{eqnarray}
|\Phi_{a'a}> &=& G_0t_a\sum_{c(\subset a')}\bar\delta_{ac}
|\Phi_{a'c}> + G_0t_aG_0(f_{a'})_a |\Phi_{a'a_1}> , \nonumber \\
|\Phi_{a'a_1}> &=& \sum_{c(\subset a')} g_0(f^\dagger_{a'})_c 
|\Phi_{a'c}> + g_0{\cal V}_{a_1}|\Phi_{a'a_1}> .
\label{wfab}
\end{eqnarray}
Because of the coupling between the $\pi$NNN and NNN sectors the
wave function $|{\bf \Phi}^{(3)}_{a'}>$ acquires an extra
component $|\Phi_{a'a_1}>$. One immediately sees that, if the coupling 
between the two spaces is switched off 
($(f_{a'})_a = (f^\dagger_{a'})_a \equiv 0$)
these equations reduce themselves to a pair of uncoupled equations,
the former having the same form as (\ref{wfags}), and describing
a deuteron $({\rm D})$ as a $\pi$NN bound state (plus a spectator 
nucleon), the latter referring to a pair of nucleons bound by the 
two--body potential ${\cal V}_{a_1}$, in presence of a spectator nucleon
in the NNN space. In presence of the $\pi$NN vertices, the total wave 
function becomes the superposition of the two components, since the 
deuteron can be viewed both as an NN bound--state, and as an NN system 
with a pion in flight between the two fermions. The states 
$|{\bf \Gamma}^{(3)}_{a'}(E_{a'})>$ are then the form factors for
the composite system in partition $a'$.

      For $a'$ belonging to the Type--II class, one gets equations
quite similar to (\ref{wfab}). The second term in the equation
for the $|\Phi_{a'a_1}>$ component, however, is now missing 
(see Eq.~(\ref{vab2})). In analogy to the standard four--body case, this
set describes an off--shell situation, with a correlated NN pair in 
presence of an interacting $\pi$N system. Since the corresponding 
form factor $|{\bf \Gamma}^{(3)}_{a'}>$ can be most clearly interpreted
in the quasi--particle scheme, we defer its discussion to the
next Section. Here, we limit ourselves to observe that Eqs.~(\ref{homo1})
or (\ref{homo2}) contain the contributions due to the dangerous self--energy 
graphs leading to the disconnectedness problems outlined in the
previous Section.

      From the formal solution of the dynamical Eqs.~(\ref{subLS}),
it follows that in momentum--space representation ${\sf t}^{(3)}_{a'}(z)$ 
has a pole for $z \sim E_{a'}$. In operator notation we simply write
\begin{equation}
{\sf t}^{(3)}_{a'}(z) \simeq 
{|{\bf \Gamma}^{(3)}_{a'}><{\bf \Gamma}^{(3)}_{a'}|\over z - E_{a'}} ,
\label{iperpole}
\end{equation}
where $<{\bf \Gamma}^{(3)}_{a'}|$ is a row vector, defined as the solution
of
\[<{\bf \Gamma}^{(3)}_{a'}(E_{a'})|=<{\bf \Gamma}^{(3)}_{a'}(E_{a'})|
{{\bf G}}^{(3)}_{0}(E_{a'}){\sf v}^{(3)}_{a'}(E_{a'}).\]

      Eq.~(\ref{iperpole}) can be given an interpretation quite similar
to the pole approximation for the sub--system amplitudes in standard 
few--body theory \cite{grass67}. For Type--I and Type--III partitions
the form factor $<{\bf \Gamma}^{(3)}_{a'}|$ describes the formation of 
a bound (or resonant) three--body subsystem in partition $a'$; the 
factor $(z-E_{a'})^{-1}$ describes the propagation of this sub--system
in presence of the spectator particle, whereas the form factor 
$|{\bf \Gamma}^{(3)}_{a'}>$ is associated to the virtual decay of the
composite object. For Type--II partitions, on the other hand, one has
the (virtual) formation, propagation and subsequent decay of two
correlated pairs with no mutual interaction. Differently from the
usual four--body problem, however, here the coupling between the
NNN and $\pi$NNN sectors introduces an extra component for the
form factors for Type-I and Type--II partitions. As already remarked,
this extra component is present because the absorption and emission
of the pion implies that the deuteron has to be regarded as a
superposition of pure NN configurations and three--body $\pi$NN
components (Type-I partitions), and the nucleon is allowed to
emit and absorb the pion in presence of a correlated NN pair
(Type-II partitions).

      Once the singular behaviour of the sub--amplitudes 
${\sf t}^{(3)}_{a'}$ has been exhibited, the physical transition 
amplitudes for two--fragment$\rightarrow$two--fragment processes 
can be extracted from Eq.~(\ref{fundamental}), by looking for the
singular components of the operator ${{\sf T}^{(3)}}$. To be 
definite, we suppose that the initial configuration is represented by 
a pion impinging on a three--nucleon system, so that the corresponding 
two--cluster partition is $b'=\pi({\rm NNN})$. We look for the transition 
amplitude to the final configuration where nucleons ${\rm N}_j$ and 
${\rm N}_k$ are bound in the deuteron, and the third nucleon ${\rm N}_i$ 
is free ($a'={\rm N}_i(\pi{\rm N}_j{\rm N}_k)$) \cite{note2}. Using
Eq.~(\ref{iperpole}) in Eq.~(\ref{fundamental}) we can exhibit
the corresponding singular part of ${{\sf T}^{(3)}}$ as follows
\begin{equation}
{{\sf T}^{(3)}} = |{\bf \Gamma}^{(3)}_{a'}>{{\cal T}_{a'b'} \over 
(z - E_{a'})(z - E_{b'})}<{\bf \Gamma}^{(3)}_{b'}| 
+ {{\sf T}}^{(3)({\rm NP})} .
\label{nddecomp}
\end{equation}
Here, ${{\sf T}}^{(3)({\rm NP})}$ is the non--singular part of 
${{\sf T}}^{(3)}$, $a'={\rm N}_i(\pi{\rm N}_j{\rm N}_k)$, 
$b'=\pi({\rm NNN})$, the form factors $|{\bf \Gamma}^{(3)}_{a'}>$
and $<{\bf \Gamma}^{(3)}_{b'}|$ satisfy the equations
\begin{eqnarray}
|\Gamma_{a'a}> &=& \sum_{c(\subset a')}\bar\delta_{ac}t_cG_0|\Gamma_{a'c}>
+ (f_{a'})_ag_0|\Gamma_{a'a_1}> \nonumber \\
|\Gamma_{a'a_1}> &=& \sum_{c(\subset a')} (f^\dagger_{a'})_cG_0t_cG_0
|\Gamma_{a'c}> + {\cal V}_{a_1}g_0|\Gamma_{a'a_1}>,
\label{deuff}
\end{eqnarray}
and
\begin{equation}
<\Gamma_{b'b}| = \sum_{c(\subset b')} <\Gamma_{b'c}|G_0t_c\bar\delta_{cb} ,
\label{triff}
\end{equation}
respectively, and ${\cal T}_{a'b'}$ is defined according to
\begin{equation}
{\cal T}_{a'b'} \equiv 
<{\bf \Phi}^{(3)}_{a'}|{{\sf U}^{(3)}_{a'b'}}|{\bf \Phi}^{(3)}_{b'}> .
\label{firstamp}
\end{equation}

      In conclusion, the two--cluster$\rightarrow$two--cluster 
transition amplitudes ${\cal T}_{a'b'}$ have been obtained as the
residue at the double pole on--shell ({\it i.e.} at $z = E_{a'} = 
E_{b'}$) of the three--cluster$\rightarrow$three--cluster amplitudes
given in Eq.~(\ref{titre}).

      The above analysis can be easily modified so as to extract the
transition amplitude for the $\pi + {\rm T} \rightarrow {\rm N} +
{\rm N} + {\rm N}$ process. Using Eqs.~(\ref{iperpole}) and
(\ref{homo2}) in Eq.~(\ref{fundamental}), and extracting the
residue corresponding to the pole at $z = E_{b'}$ 
($b' = \pi({\rm N}{\rm N}{\rm N})$) one gets from
$<\chi_{0}'|U^\dagger_b|\phi_b>$ the following expression for the
transition amplitude ${\cal T}_{{\rm NNN}\leftarrow \pi{\rm T}}$
\begin{eqnarray}
{\cal T}_{{\rm NNN}\leftarrow \pi{\rm T}} &=&
\nonumber \\
&=& \sum_{a' (\in {\rm I,II})}\sum_b
\{\sum_a <\chi_{0}'|(u^\dagger_{a'})_a G_0t_aG_0 U_{a'ab'b}|\Phi_{b'b}>
+ <\chi_{0}'|u_{a'}g_0U^\dagger_{a'a_1 b'b}|\Phi_{b'b}>\} ,
\label{thirdamp}
\end{eqnarray}
where the sum over $a'$ is restricted to Type--I and Type--II partitions 
only, and we have implicitly assumed $a\subset a'$. Thus, in close analogy 
to what is done in ordinary scattering theory, by picking up the appropriate 
singular part of the three--cluster$\rightarrow$ three--cluster transition 
operators, we have identified the contribution corresponding, in 
configuration space, to an incoming wave with a pion and 
a bound three--nucleon system, plus a scattered wave with three free 
nucleons in the asymptotic region.

      We also report the expressions for the partial-- and 
total--break--up transition amplitudes. With a self--explanatory
notation one can write
\begin{eqnarray}
{\cal T}_{\pi {\rm N}{\rm D}\leftarrow \pi{\rm T}} &=&
\sum_{a'}\sum_{c(\subset a')}\sum_{b}
<\phi_a|(u_{a'})_{ac}G_0t_cG_0U_{a'cb'b}|\Phi_{b'b}>
\nonumber \\
&+& \sum_{a'(\in {\rm I,II})}\sum_b <\phi_a|(u_{a'})_a
g_0U^\dagger_{a'a_1b'b}|\Phi_{b'b}> ,
\label{fouramp}
\end{eqnarray}
where $<\phi_a|$ is the three--cluster asymptotic state with a deuteron
plus spectators nucleon and pion (see Eq.(~\ref{chaneq})), and
\begin{eqnarray}
{\cal T}_{\pi {\rm NNN}\leftarrow \pi{\rm T}} &=& \sum_{a'}\sum_a
\sum_{c(\subset a')}\sum_b 
<\chi_{1}'|t_aG_0(u_{a'})_{ac}G_0t_cG_0U_{a'cb'b}|\Phi_{b'b}>
\nonumber \\
&+& \sum_{a'(\in {\rm I,II})}\sum_a\sum_b <\chi_1'|t_aG_0(u_{a'})_a
g_0U^\dagger_{a'a_1b'b}|\Phi_{b'b}> .
\label{fiveamp}
\end{eqnarray}
Here, obviously, $b'=\pi({\rm NNN})$, and $a$ in Eq.~(\ref{fouramp}) 
identifies the partition $\pi{\rm N}_i({\rm N}_j{\rm N}_k)$.

      For the process $\pi + {\rm T} \rightarrow {\rm N} + {\rm D}$, the 
identification of ${\cal T}_{a'b'}$, as given by Eq.~(\ref{firstamp}),
with the required transition amplitude 
${\cal T}_{{\rm N}{\rm D}\leftarrow \pi {\rm T}}$ 
deserves some more comments. In the first place, it is worth to note that, 
since only states with at most one pion are explicitly taken into account,
the triton and the deuteron are treated in a different way in the
respective channels. The former is described as a three--nucleon system
bound by static NN forces because of the presence of the real pion
(see Eqs.~(\ref{wfags}) or (\ref{triff})), the latter as a superposition 
of NN and $\pi$NN configurations (see Eqs.~(\ref{wfab}) or (\ref{deuff})).

      It may be instructive to write Eq.~(\ref{firstamp}) explicitly. 
Recalling that $|{\bf \Phi}^{(3)}_{b'}>$ has no hybrid--chain component 
for $b'$ of Type-III, one gets
\begin{equation}
{\cal T}_{{\rm N}{\rm D}\leftarrow \pi {\rm T}} = 
\sum_{ab}<\Phi_{a'a}|U_{a'a b'b}|\Phi_{b'b}> + \sum_b
<\Phi_{a'a_1}|U^\dagger_{a'a_1 b'b}|\Phi_{b'b}> .
\label{firstampex}
\end{equation}
The first term in this expression has the same form as the corresponding
transition amplitude in the standard GS formulation of the four--body
problem \cite{san80,cv83}, the channel states and the transition
operators satisfying now equations allowing for the coupling between
the NNN and the $\pi$NNN spaces. The latter term explicitly takes
into account the possible transition between the two spaces, when
the pion is absorbed.

      It is worth to observe that $2+2$ self--energy graphs
imply non--trivial difficulties also for the transition amplitudes.
Indeed, in the present framework one has to distinguish between 
${\rm N}(\pi{\rm NN})$ and $(\pi{\rm N})({\rm N}{\rm N})$ partitions. 
As we shall see in the next Section, this is consistent with a 
quasi--particle scheme, where the physical transition amplitudes 
are evaluated in terms of two-- and three--body quasiparticles. Under
the point of view of a full, renormalized field--theory of interacting
pions and nucleons, however, Type--I and Type--II partitions
have to concur to describe the same physical situation, a 
dressed nucleon with its physical mass, which asymptotically
propagates in presence of a deuteron, bound by the exchange of
pions between the constituent nucleons. To meet these physical
requirements the interactions should be constrained so as to guarantee
that the sub--system operators ${{\sf t}^{(3)}_{a'}}$ have a 
pole at the same energy $E_{a'}$ for both Type--I and Type--II
partitions. For $\pi+ {\rm T} \rightarrow {\rm N} + {\rm D}$
scattering Eq.~(\ref{firstamp}) would be then replaced by
\begin{equation}
{\cal T}_{{\rm N}{\rm D}\leftarrow \pi {\rm T}} = 
\sum_{a'(\in {\rm I,II})} 
<{\bf \Phi}^{(3)}_{a'}|{{\sf U}^{(3)}_{a'b'}}|{\bf \Phi}^{(3)}_{b'}>,
\label{correctamp}
\end{equation}
where (apart from anti--symmetrization) one has 
$a'={\rm N}_i(\pi {\rm N}_j{\rm N}_k)$ and
$a'=(\pi {\rm N}_i)({\rm N}_j{\rm N}_k)$. That such a description cannot
emerge in a natural way from the present theory is a consequence
of the truncation to states with at most one pion, which lies at the
heart of this type of approaches. This difficulty, which is somewhat
hidden in the $\pi$NN--NN case (apart from the renormalization
problem), shows up here owing to the more complex boundary conditions.
If the coupling between the $\pi$NNN and NNN sectors is switched off
for Type--II partitions, on the other hand, this ambiguity is removed;
Eqs.~(\ref{homo2}) reduce to the usual AGS Eqs.~(\ref{wfags}) for
two pairs of correlated particles, and one can distinguish between
the physical, spectator nucleon in Type-I partitions, and the
correlated $\pi$N pair in the Type-II ones. As a consequence, 
Eq.~(\ref{firstamp}) yields now the correct transition amplitude
without ambiguities. Thus, the same truncation which leads to a 
connected--kernel scheme, allows one to unambiguously identify the 
physical transition amplitudes for two--cluster$\rightarrow$ 
two--cluster processes.

\section{The Quasi--Particle Scheme}
\label{qps}

      The quasi--particle approximation (QPA) has played a major role 
in the development of conventional few--body scattering theory 
\cite{grass67,ak91}. This approach relies on the property that both the 
two--body t--matrices and the sub--system amplitudes can be approximated 
to an arbitrary degree of accuracy by a sum of separable terms. A first
application of the QPA to the two--body t--operators allows one to
re--write the original four--body YGS equations as Faddeev-type equations 
for two elementary particles and a composite object. A second 
application of the QPA to the sub--system amplitudes reduces the
dynamical equations to a set of effective Lippmann--Schwinger equations 
coupling the two--cluster$\rightarrow$two--cluster transition amplitudes. 
These equations, after partial--wave projection, involve one integration 
variable only. The break--up amplitudes can be then obtained from the
2$\rightarrow$2 amplitudes by quadrature. The QPA allows for a 
better insight into the physical content of Yakubovski\u{\i}--type 
formalisms, at the same time alleviating the difficulties implied by the 
solution of the four--body scattering problem by brute force.
\par
      Here, we show that the QPA can be naturally extended to the
present situation, where emission/absorption processes are allowed.
We shall assume a separable (rank one) form for the two--body t--operators 
as well as for the sub--system amplitudes, the extension to the general 
(finite--rank) case being straightforward. Thus, we write 
\begin{equation}
t_a(z) = |a>\tau^{(3)}_a(z)<a| ,
\label{sept}
\end{equation}
with $|a> = V_a|\phi_a>$ and $\tau^{(3)}_a(z) \sim 1/(z-E_a)$ for $z \simeq 
E_a$. The actual form of $\tau^{(3)}_a$ depends on the particular 
separable--expansion method employed. However, all the reasonable 
expansion methods must reproduce the bound--state pole behaviour of 
Eq.~(\ref{pole}) \cite{lcr}. Then, the dynamical equations (\ref{subLS}) 
for the sub--system amplitudes can be re--written as
\begin{equation}
{\sf X}^{(3)}_{a'}={\sf Z}^{(3)}_{a'}+
{\sf Z}^{(3)}_{a'}{\sf D}^{(3)}{\sf X}^{(3)}_{a'} ,  
\label{subLSsep}
\end{equation}
where ${\sf D}^{(3)}$, ${\sf Z}^{(3)}_{a'}$, and 
${\sf X}^{(3)}_{a'}$ are again matrices in the 
three--cluster--partition indices, having a block structure quite 
similar to ${\sf G}_0^{(3)}$, ${\sf v}^{(3)}_{a'}$ and 
${\sf t}^{(3)}_{a'}$. Thus, one has
\begin{mathletters}
\begin{equation}
{\sf D}^{(3)}
\equiv 
\left|       
\begin{array}{cc}
{D^{(3)}}_{(a|b)} & {D^{(3)}}_{(a|0)} \cr 
{D^{(3)}}_{(0|b)} & {D^{(3)}}_{(0|0)} \cr
\end{array} \right| 
= 
\left| \begin{array}{ccc}
\tau^{(3)}_a\delta_{ab} &\ & 0 \\
0 &\ & g_0 \\
\end{array} \right| .
\label{propsep}
\end{equation}
      Similarly, for Type--I partitions, ${\sf Z}_{a'}^{(3)}$ and 
${\sf X}_{a'}^{(3)}$ are defined according to
\begin{eqnarray}
{\sf Z}_{a'}^{(3)}
\equiv 
&\left| 
\begin{array}{cc}
{Z_{a'}^{(3)}}_{(a|b)} & {Z_{a'}^{(3)}}_{(a|0)} \cr 
{Z_{a'}^{(3)}}_{(0|b)} & {Z_{a'}^{(3)}}_{(0|0)} \cr
\end{array} \right| 
&= 
\left| \begin{array}{cc}
<a|G_0|b>\bar\delta_{ab}\delta_{a,b\subset a'} & <a|G_0(f_{a'})_{a}  \\
         (f^\dagger_{a'})_{b}G_0|b>            &   {\cal V}_{a_1}    \\
\end{array} \right| ,
\label{vabsep1} \\
{\sf X}_{a'}^{(3)}
\equiv 
& \left| 
\begin{array}{cc}
{X_{a'}^{(3)}}_{(a|b)} & {X_{a'}^{(3)}}_{(a|0)} \cr 
{X_{a'}^{(3)}}_{(0|b)} & {X_{a'}^{(3)}}_{(0|0)} \cr
\end{array} \right| 
&= 
\left| \begin{array}{ccc}
<a|G_0(u_{a'})_{ab}G_0|b> &\ & <a|G_0(u_{a'})_{a}\\
(u^\dagger_{a'})_bG_0|b>  &\ &        u_{a'}     \\
\end{array} \right| .
\label{subsep} 
\end{eqnarray}
\end{mathletters}
For $a'$ of Type--II one has similar definitions, with the 
element ${Z_{a'}^{(3)}}_{(0|0)}$ missing,
whereas, for the Type--III partition $\pi$(NNN), only the upper left
blocks survive, thereby reproducing the definition of standard
four--body theory.

      The meaning of the present notation is worth of some comment.
As in the four--body GS approach, in $<a|G_0|b>$ an
integration has been performed over the relative momenta of the
correlated pairs described by the form factors $<a|$ and $|b>$,
so that this quantity is still an operator with respect to the 
remaining spectator momenta. In momentum--space
representation and for Type-I and Type--III partitions 
${Z_{a'}^{(3)}}_{(a|b)}$ describes the virtual decay of the correlated 
pair $b$ and the formation of a correlated pair $a$, with the 
intermediate exchange of a particle between the two fragments. These
$Z$--terms play therefore the role of intercluster interactions,
originated from the exchange of a particle. For Type--II partitions one 
has the virtual decay of $b$ and the subsequent formation of $a$, with 
intermediate four--body propagation. 

      Let's now consider $<a|G_0(f_{a'})_{a}$. In this case we integrate 
over the relative momentum of pair $a$ in the $\pi$NNN sector only; such
a quantity describes a transition from the NNN space to the $\pi$NNN 
space, with formation of a correlated pair $a$, describing either a 
bound deuteron or a resonant $\pi$N system. A graphical 
interpretation of $<a|G_0(f_{a'})_{a}$ is given in Fig.~\ref{fig2}. 
Similar considerations apply to $<a|G_0(u_{a'})_{a}$ and 
$(u^\dagger_{a'})_{b}G_0|b>$ ($b\subset a'$). In summary, we are 
consistently applying the GS recipe in the four--body sector, the NNN 
space remaining untouched, so as to get operators in two vector variables 
in both sectors.

      From the Eqs.~(\ref{yaku}) the set of equations
\begin{equation}
{\sf Y}^{(3)}_{a'b'} =\bar\delta_{a'b'}{{\sf D}^{(3)}}^{-1}+
\sum_{c'}\bar\delta_{a'c'}{\sf X}^{(3)}_{c'}
{\sf D}^{(3)}{\sf Y}^{(3)}_{c'b'} ,
\label{yakusep}
\end{equation}
can be derived, where
\begin{equation}
{\sf Y}^{(3)}_{a'b'} \equiv
\left |
\begin{array}{cc}
Y^{(3)}_{a'ab'b}   & Y^{(3)}_{a'ab'b_1}   \cr
Y^{(3)}_{a'a_1b'b} & Y^{(3)}_{a'a_1b'b_1} \cr
\end{array} \right |
=
\left |
\begin{array}{ccc}
<a|G_0U_{a'a b'b}G_0|b>     &\   & <a|G_0U_{a'a b'b_1} \cr
U^\dagger_{a'a_1 b'b}G_0|b> &\   & U_{a'a_1 b'b_1}    \cr
\end{array} \right | .
\label{yakpos}
\end{equation}

      To introduce separable approximations for the sub--system
amplitudes we observe that the homogeneous equations (\ref{homo2})
become, when the two--body t-matrices are approximated by 
Eq.~(\ref{sept}),
\begin{equation}
|\tilde {\bf \Phi}^{(3)}_{a'}(E_{a'})> = {\sf D}^{(3)}(E_{a'})
{\sf Z}^{(3)}_{a'}(E_{a'})|\tilde {\bf \Phi}^{(3)}_{a'}(E_{a'})> ,
\label{homo2sep}
\end{equation}
with $|\tilde {\bf \Phi}^{(3)}_{a'}(E_{a'})>$ a column vector, whose 
components are related to the components of 
$|{\bf \Phi}^{(3)}_{a'}(E_{a'})>$ by the remarkable factorization relation
\begin{equation}
|\Phi_{a'a}> = G_0|a>|\tilde \Phi_{a'a}> ,\;\; 
|\Phi_{a'a_1}> \equiv |\tilde \Phi_{a'a_1}> .
\label{factor}
\end{equation}
Needless to say, the NNN--components of these vectors are missing
for the Type-III partition. The form--factors $|{\bf \Gamma}^{(3)}_{a'}>$ 
are now replaced by the vectors $|\tilde {\bf \Gamma}^{(3)}_{a'}> = 
{\sf Z}^{(3)}_{a'}|\tilde {\bf \Phi}^{(3)}_{a'}>$
satisfying the analogue of Eqs.~(\ref{homo1})
\begin{equation}
|\tilde {\bf \Gamma}^{(3)}_{a'}(E_{a'})> = {\sf Z}^{(3)}_{a'}(E_{a'})
{\sf D}^{(3)}(E_{a'})|\tilde {\bf \Gamma}^{(3)}_{a'}(E_{a'})> .
\label{homo1sep}
\end{equation}
      The physical meaning of the ``form factors" 
$|\tilde {\bf \Gamma}^{(3)}_{a'}>$ is the following. 
For $a'=\pi ({\rm NNN})$, 
$|\tilde \Gamma_{a'a}>$ describes the virtual decay of the three--nucleon 
bound state into the correlated pair $a$ plus a nucleon, in presence of 
the spectator pion. For the Type--III partition, therefore, the three 
components $|\tilde \Gamma_{a'a}>$ retain the same meaning as in the 
standard four--body problem \cite{grass67}, and satisfy the homogeneous 
Faddeev--type equations
\begin{equation}
|\tilde \Gamma_{a'a}> = \sum_{c(\subset a')}\bar\delta_{ac}<a|G_0|c>
\tau^{(3)}_c |\tilde \Gamma_{a'c}> .
\label{ffe1}
\end{equation}
      For Type--I partitions, the form factors 
$|\tilde {\bf \Gamma}^{(3)}_{a'}>$ acquire an extra component 
$|\tilde \Gamma_{a'a_1}>$ in the three--nucleon sector. Again this is 
due to the presence of the $\pi$NN vertex functions; whereas 
$|\tilde \Gamma_{a'a}>$ represent the virtual decay of the deuteron as a 
$\pi {\rm NN}$ bound state into a 
correlated (NN) or ($\pi$N) pair plus a free particle, 
$|\tilde \Gamma_{a'a_1}>$ describes the decay of the deuteron into a
pair of nucleons, through internal absorption of the pion. The four
components of $|\tilde {\bf \Gamma}^{(3)}_{a'}>$ are coupled by the 
Eqs.~(\ref{homo1sep}), which explicitly written give
\begin{eqnarray}
|\tilde \Gamma_{a'a}> &=& \sum_{c(\subset a')}\bar\delta_{ac}<a|G_0|c>
\tau^{(3)}_c |\tilde \Gamma_{a'c}> + 
<a|G_0(f_{a'})_ag_0|\tilde \Gamma_{a'a_1}> \nonumber \\
|\tilde \Gamma_{a'a_1}> &=& \sum_{c(\subset a')}(f^\dagger_{a'})_c G_0|c>
\tau^{(3)}_c |\tilde \Gamma_{a'c}> + 
{\cal V}_{a_1} g_0|\tilde \Gamma_{a'a_1}> .
\label{ffe2}
\end{eqnarray}
These equations are graphically illustrated in Fig.~\ref{fig3}. Finally,
the form--factor components $|\tilde \Gamma_{a'a}>$ corresponding to 
Type--II partitions describe in the $\pi$NNN sector off--shell situations
in which a deuteron or a correlated $\pi$N pair decays in presence of another
correlated pair. The ``absorption" component $|\tilde \Gamma_{a'a_1}>$,
on the other hand, is associated to a deuteron which decays while a
($\pi$N) quasi--particle undergoes an (off--shell) 
$(\pi{\rm N})\rightarrow$N transition. These form factors and their coupling 
through Eqs.~(\ref{homo1sep}) are graphically depicted in Fig.~\ref{fig4}.

      Consistently with the discussion of the residue method given
in Section \ref{res}, we finally assume for ${\sf X}^{(3)}_{a'}$
the separable representation
\begin{equation}
{\sf X}^{(3)}_{a'} = |\tilde {\bf \Gamma}^{(3)}_{a'}>
\tau^{(2)}_{a'} <\tilde {\bf \Gamma}^{(3)}_{a'}| ,
\label{iperpolesep}
\end{equation}
with $\tau^{(2)}_{a'} \simeq 1/(z-E_{a'})$ for $z\sim E_{a'}$, and
the row vector $<\tilde {\bf \Gamma}^{(3)}_{a'}|$ eigensolution
of \break $<\tilde {\bf \Gamma}^{(3)}_{a'}|=<\tilde {\bf \Gamma}^{(3)}_{a'}|
{\sf D}^{(3)} {\sf Z}^{(3)}_{a'}$.

      Inserting (\ref{iperpolesep}) into the Eqs.~(\ref{yakusep}),
and evaluating the resulting equations between \break
$<\tilde {\bf \Gamma}^{(3)}_{a'}|{\sf D}^{(3)}$ on the left
and ${\sf D}^{(3)}|\tilde {\bf \Gamma}^{(3)}_{b'}>$ on the right
one gets 
\begin{equation}
{\cal T}_{a'b'} = \bar\delta_{a'b'}<\tilde {\bf \Gamma}^{(3)}_{a'}|
{\sf D}^{(3)}|\tilde {\bf \Gamma}^{(3)}_{b'}> + 
\sum_{c'}\bar\delta_{a'c'}<\tilde {\bf \Gamma}^{(3)}_{a'}|
{\sf D}^{(3)}|\tilde {\bf \Gamma}^{(3)}_{c'}>\tau_{c'}^{(2)}
{\cal T}_{c'b'} ,
\label{superlssep}
\end{equation}
where use has been made of the relation between the form factors
$|\tilde {\bf \Gamma}^{(3)}_{a'}>$ and the ``channel states" 
$|\tilde {\bf \Phi}^{(3)}_{a'}>$, and we have now defined 
${\cal T}_{a'b'}$ according to
\begin{equation}
{\cal T}_{a'b'} \equiv <\tilde {\bf \Phi}^{(3)}_{a'}|{\sf Y}^{(3)}_{a'b'}
|\tilde {\bf \Phi}^{(3)}_{b'}> .
\label{newamp}
\end{equation}
That Eq.~(\ref{newamp}) actually gives (on shell) the 
2--cluster$\rightarrow$2--cluster transition amplitudes
in the separable approximation  can be easily ascertained starting
from Eq.~(\ref{firstamp}), and using the factorization property 
(\ref{factor}) and Eq.~(\ref{yakpos}).
Much as in the standard four--body problem, we have been 
able to obtain one--vector--variable integral equations
coupling all the 2$\rightarrow$2 transition amplitudes one has
for a given initial configuration. By means of an obvious
matrix notation in the two--cluster--partition indices 
Eq.~(\ref{superlssep}) can be re--written as an effective 
multi--channel LS equation:
\begin{equation}
{\sf X}^{(2)}={\sf Z}^{(2)}+{\sf Z}^{(2)}{\sf D}^{(2)}{\sf X}^{(2)} ,
\label{finalsuper}
\end{equation}
where
\begin{mathletters}
\begin{equation}
{\sf X}^{(2)}_{a'b'}  \equiv  {\cal T}_{a'b'} ,
\end{equation}
\begin{equation}
{\sf Z}^{(2)}_{a'b'}  \equiv  \bar\delta_{a'b'}
<\tilde {\bf \Gamma}^{(3)}_{a'}|{\sf D}^{(3)}
|\tilde {\bf \Gamma}^{(3)}_{b'}> ,
\label{driving}
\end{equation}
\begin{equation}
{\sf D}^{(2)}_{a'b'} \equiv \tau^{(2)}_{a'}\delta_{a'b'} .
\end{equation}
\end{mathletters}

      As is typical of the quasi--particle approach, the operator 
$\tau^{(2)}_{a'}$ plays the role of an effective propagator; for
Type--I partitions it describes a deuteron (regarded as a coupled 
NN--$\pi$NN system) in presence of a spectator nucleon; for Type--II 
partitions it is associated to a deuteron (as a bound NN system) 
propagating in presence of a $\pi$N correlated pair; for 
the Type--III partition it simply describes the NNN bound state and 
the non--interacting pion in the two--cluster intermediate state. 
The driving terms ${\sf Z}^{(2)}_{a'b'}$ represent exchange ``potentials" 
associated to the simplest reaction mechanisms through which one can
pass from the initial configuration $b'$ to the final one $a'$.
Owing to the $\pi$NN vertices, they contain further contributions
with respect to standard four--body theory. This can be immediately
seen by writing them explicitly as
\begin{equation}
{\sf Z}^{(2)}_{a'b'} = \bar\delta_{a'b'}\left\{\sum_{c(\subset a',b')} 
<\tilde \Gamma_{a'c}|\tau^{(3)}_{c}|\tilde \Gamma_{b'c}> +
<\tilde \Gamma_{a'a_1}|g_0|\tilde \Gamma_{b'b_1}> \right\},
\label{expexpot}
\end{equation}
where it is assumed that the latter term on the right--hand--side is
missing if either $a'$ or $b'$ is the Type--III partition. In 
particular, in standard four--body theory the driving term 
${\sf Z}^{(2)}_{a'b'}$ is missing when both $a'$ and $b'$ 
are of Type II, since in such a case no three--cluster partition 
$c$ can be simultaneously contained in both $a'$ and $b'$ (with
$a' \not= b'$). Here, this driving term survives, due to the
presence of the latter term in Eq.~(\ref{expexpot}). Some typical
driving terms are graphically illustrated in Fig.~\ref{fig5}.

      Let us focus our attention on Fig. \ref{fig5}(d). It describes one
of the lowest--order contributions to the transition ($\pi$N) + (NN)
$\rightarrow$ N + (NN$\pi$), and consists of two graphs,
corresponding to the former and the latter term on the right of
Eq.~(\ref{expexpot}), respectively. The graph associated to
intermediate three--nucleon propagation is clearly disconnected.
This is at variance with standard four--body theory, where all the
driving terms are connected after the repeated application of the
QPA, and reflects the presence of the disconnected terms in 
the kernel of the exact Eqs.~(\ref{yaku}) we have already pointed 
out in Section \ref{revrev}. One can easily convince oneself that the 
disconnected contributions due to the graph of Fig.~\ref{fig5}(d) 
survive after any number of iterations of the effective LS 
Eqs.~(\ref{finalsuper}). The above disconnected graph can
be exhibited as self--energy insertions in Type--II partitions by 
substituting the Eq.~(\ref{homo1sep}) (or an iteration of the same 
equation) into the expression (\ref{driving}) for the driving terms. 
Due to the employment of the QPA, all the disconnected terms in the 
present formalism appear now lumped together in a unique contribution, 
describing transitions from a Type--II to a Type--I partition,
having an NN pair in common. Clearly, if this term is disregarded,
one has the minimal truncation able to give connected--kernel
equations for the 2$\rightarrow$2 transition amplitudes.
If, on the other hand, one switches off completely the coupling 
between the $\pi$NNN and NNN sectors in Type--II partitions, the 
Eqs.~(\ref{homo1sep}) for the $2+2$ quasi--particle form factors 
reduce to the usual AGS homogeneous equations (\ref{ffe1}). These
are graphically depicted in Fig.~\ref{fig4}, where the boxes
enclose the terms which are absent in this approximation.
The effective LS equations (\ref{finalsuper}) remain formally
unchanged, but now in $2+2$ intermediate states one has the
propagation of a ($\pi$N) ``cluster", with no ($\pi$N)$\rightarrow$N
transition. Correspondingly, in the driving terms the contributions
enclosed in boxes in Fig. \ref{fig5} vanish; in particular,
the 2$\rightarrow$2 graph is now absent, much as in standard
four--body theory.

      Finally, let us briefly describe how the above considerations
have to be modified, if $\pi$ absorption/emission is taken
into account in the scattering amplitudes for $2+2$ sub--systems,
as described by Eqs.~(\ref{solveq}). The quasi--particle representation
for these equations still has an LS--type structure 
\begin{equation}
{\sf X}^{(3)}_{a'} = {\sf Z}^{(3)}_{a'} + {\sf Z}^{(3)}_{a'}
{\sf D}^{(3)}{\sf X}^{(3)}_{a'} ,
\label{newLS}
\end{equation}
where ${\sf D}^{(3)}$  and ${\sf X}^{(3)}_{a'}$  can be identified
with $D^{(3)}_{(a|b)} = \tau^{(3)}_a \delta_{ab}$ and 
${X^{(3)}_{a'}}_{(a|b)} = <a|G_0(u_{a'})_{ab}G_0|b>$, respectively,
and the $Z$--term ${\sf Z}^{(3)}_{a'}$ contains two terms,
the former being $<a|G_0|b>~\bar\delta_{ab}~\delta_{a,b\subset a'}$,
the latter, given by $<a|G_0 (f_{a'})_a g_0 (f^\dagger_{a'})_b G_0|b>$,
containing the effects of $\pi$ emission/absorption in the
scattering channel. This contribution is graphically depicted
in Fig.~\ref{fig6}.

	    To conclude this Section, we give the physical transition 
amplitudes for absorption/emission and break--up processes in terms of 
the quasi--particle amplitudes ${\cal T}_{a'b'}$. Inserting 
(\ref{iperpolesep}) into Eq.~(\ref{thirdamp}) and using (\ref{newamp}) 
one has for the $\pi + {\rm T} \rightarrow {\rm N}{\rm N}{\rm N}$
process 
\begin{equation}
{\cal T}_{{\rm NNN}\leftarrow \pi {\rm T}} =
\sum_{a'} <\chi_{0}'|\tilde \Gamma_{a'a_1}>
\tau^{(2)}_{a'}{\cal T}_{a' b'} ,
\label{absampsep}
\end{equation}
with $b' = \pi ({\rm NNN})$. This result is what one could reasonably 
expect in the framework of an isobar model; the absorption amplitude 
with three free outgoing nucleons is given as the sum over all possible 
transitions to two--body intermediate states, followed by the decay of 
the composite clusters or virtual $(\pi {\rm N}) \rightarrow {\rm N}$ 
conversion. This is graphically shown in Fig. \ref{fig7}. Similarly, 
one has for the break--up transition amplitudes 
\begin{equation}
{\cal T}_{\pi{\rm N}{\rm D}\leftarrow \pi {\rm T}} =
\sum_{a'(\supset a)}<\phi_a|\tilde \Gamma_{a'a}>
\tau^{(2)}_{a'}{\cal T}_{a'b'} \qquad (a\equiv \pi N D) ,
\label{parampsep}
\end{equation}
\begin{equation}
{\cal T}_{\pi {\rm NNN}\leftarrow \pi {\rm T}} =
\sum_{a'a}<\chi_1'|a>\tau^{(3)}_{a}<\phi_a|\tilde \Gamma_{a'a}>
\tau^{(2)}_{a'}{\cal T}_{a'b'} .
\label{totampsep}
\end{equation}

\section{Conclusions and comparison with previous approaches}
\label{finis}

      In this paper we have revisited the approach to the $\pi$NNN--NNN
problem presented in Refs. \cite{cc1,cc2}. This formalism can be
regarded as the most natural generalization of the GS method, to a
situation with a variable number of particles. We have shown that
the physical transition amplitudes can be related to the 
chain--of--partition labelled operators of the formalism {\it via}
the residue method, and that the quasi--particle technique can be 
generalized, so as to get multichannel LS equations coupling all the 
relevant 2$\rightarrow$2 transition amplitudes. We have seen, however,
that, in spite of its attractive features, the present approach fails
in producing connected--kernel equations. This is due to off--shell
self--energy graphs, where nucleon dressing occurs in presence of
an interacting NN pair. These disconnected graphs contribute to
one single term in the kernel of the effective LS
equations for the transition amplitudes. If this term is neglected,
one is left with well--behaved equations, at the price of
sacrificing exact unitarity. A more severe truncation consists
in switching off the absorption/emission vertices in 2+2 partitions,
while keeping the effects of pion emission/absorption in the 
$\pi$NNN sub--system. At this point, a comparison with the treatment
of the $\pi$NNN--NNN problem proposed in Ref. \cite{avi83} is
suitable. There, one assumes the separable approximation (\ref{sept})
for the two--body t--operators from the beginning, and applies the
quasi--particle method directly to the AB Eqs.~(\ref{lsab}). By
the same procedure which led us to Eqs.~(\ref{subLSsep}) one gets
\begin{equation}
{\sf X}^{(3)} = {\sf Z}^{(3)} + {\sf Z}^{(3)}{\sf D}^{(3)}{\sf X}^{(3)} ,
\label{lsavi}
\end{equation}
with ${\sf D}^{(3)}$ given by Eq.~(\ref{propsep}), and
\begin{equation}
{\sf Z}^{(3)}
\equiv 
\left | 
\begin{array}{cc}
<a|G_0|b>\bar\delta_{ab}   &   <a|G_0F_a  \\
F^\dagger_b G_0|b>         &   {\cal V}   \\
\end{array} \right |
\equiv
\left |
\begin{array}{ccc}
{\sf V} & \ & {\sf q}  \\
{\sf p} & \ & {\cal V} \\
\end{array} \right |  ,
\label{zetaavi}
\end{equation}
\begin{equation}
{\sf X}^{(3)}
\equiv 
\left | 
\begin{array}{cc}
   <a|G_0U_{ab}G_0|b>    &   <a|G_0U_a  \\
   U^\dagger_b G_0|b>    &        U     \\
\end{array} \right | .
\label{xavi}
\end{equation}
      Here, according to the notation of Ref. \cite{avi83}, ${\sf V}$,
${\sf p}$, and ${\sf q}$ are a square matrix, a row-- and a 
column--vector, respectively, in the three--cluster partition
indices. Notice that, owing to the sum rules \cite{cc1,cc2}
\begin{equation}
F_a = \sum_{a' (\in {\rm I, II})} (f_{a'})_a\qquad\qquad
F^\dagger_a = \sum_{a' (\in {\rm I, II})} (f^\dagger_{a'})_a ,
\label{sumcit}
\end{equation}
one can immediately decompose ${\sf q}$ and ${\sf p}$ into terms
of well--defined clustering nature,
\begin{equation}
{\sf q} = \sum_{a' (\in {\rm I, II})} {\sf q}_{a'} \qquad\qquad
{\sf p} = \sum_{a' (\in {\rm I, II})} {\sf p}_{a'} .
\label{sumcoup}
\end{equation}

      The integral Eqs.~(\ref{lsavi}) are transformed into a set
of equivalent Schr\"odinger equations, coupling the $\pi$NNN and
NNN sectors, and the usual GS approach is employed first, by
decomposing ${\sf V}$ into terms labelled by two--cluster partitions,
so as to remove the disconnected terms one has in the standard
four--body problem. The transition operators for the various channels
of the $\pi$NNN--NNN system are then identified, and corresponding
sets of integral equations are derived. The kernel of these equations
still contains, through the coupling operators ${\sf q}$ and ${\sf p}$,
all the disconnected contributions due to emission/absorption processes.
These disconnected pieces are identified, in virtue of the 
decompositions (\ref{sumcoup}), and removed by formal use of the
two--potential formula. As a consequence, the physical transition
operators are given by the solution of rather complicate sets of
nested integral equations. This procedure implies also that the
transition operators for emission/absorption processes have to be
defined with reference to incomplete NN interactions in the NNN
space, taking account of the coupling to the $\pi$NNN sector only
to low order (see Eqs. (4.16b), (4.16c) and (4.17a) in 
Ref. \cite{avi83}). Only the exact employment of the two--potential
formula can introduce the missing part of the NN interactions,
and provide at the end the physical transition amplitudes.

      Coupled equations formally similar to the LS 
Eqs.~(\ref{finalsuper}) represent the starting point of the
relativistic, coupled--channel approach of Ref. \cite{ueda}.
There, however, the four--body dynamics never comes explicitly
into play; rather, the various three--cluster partitions for the
$\pi$NNN--NNN system are regarded as different three--body problems,
and the corresponding two--cluster partitions as different channels
of the same effective, coupled--channel problem. Thus, regarding
nucleons as distinguishable, one has nine different channels
$({\rm N}_i {\rm D}_{jk})\pi$, $(\pi {\rm D}_{jk}){\rm N}_i$, and 
$(\pi {\rm N}_i){\rm D}_{jk}$ ($i$, $j$, $k$ a cyclic permutation of
1, 2, 3). Similarly, one has nine channels for the $\pi$N$\Delta$
systems, plus six possible channels in the NNN space. Overall,
one has to deal with 24 different channels, which are coupled
through driving terms $Z_{\alpha \beta}$, describing standard
AGS exchange graphs. The various three--cluster systems are
moreover coupled because the inter--cluster interactions are
regarded as effective coupled--channel potentials; for instance, the
$\pi {\rm D}_{23}$--N$_2\Delta$--N$_3\Delta$--N$_2$N$_3$ systems
are described through a four--channel interaction. Assuming
a finite--rank form for these potentials, one has effective
two--body propagators in the assumed dynamical equations. Since
the Blankenbecler--Sugar \cite{bs66} and Aaron--Amado--Young
\cite{aay} propagators are used for the two-- and three--body
sub--systems, unitarity is satisfied up to the three--body level.
Clearly, owing to the above features, this formalism cannot be
founded on rigorous, FY--type approaches; in particular, it
implies a direct coupling between the $({\rm N}{\rm D})\pi$ configuration
({\it i. e.} the Type-III partition) in the $\pi$NNN space, and
the ${\rm N}{\rm D}$ system in the NNN sector, whereas, according to
microscopic four--body approaches, absorption has to be excluded
in the partition where the $\pi$ has a passive role in the
sub--system dynamics \cite{cc1,cc2,avi83}.

      In conclusion, to the best of our knowledge, the coupled 
$\pi$NNN--NNN problem has not found up to now a fully satisfactory
solution. The GS approach introduced in \cite{cc1,cc2} can provide
connected--kernel equations only when a class of disconnected
graphs with self--energy insertions is excluded, thereby violating
unitarity. These disconnected contributions can be removed through 
the two--potential formulation of Ref. \cite{avi83}; the physical
transition amplitudes, however, can be evaluated only through the
solution of cumbersome sets of nested integral equations. The
effective, coupled--channel method of Ref. \cite{ueda}, while allowing
actual computations of $\pi$NNN--NNN processes, relies on several
{\it ad hoc} assumptions, and cannot be directly connected to
what is suggested by rigorous FY--type analyses.

\begin{figure}
\caption{(a) Self--energy graphs occurring in Type--II partitions.
The full and dashed lines represent nucleons and pions, respectively,
while the full circles represent the $\pi$NN vertices, and the blob is
the NN t--matrix; (b) a disconnected contribution to the third
iteration of the kernel ${\sf K}$. The wavy line 
represents the NN potential in the three--nucleon space.
\label{fig1}}
\end{figure}
\begin{figure}
\caption{Graphical representation of $<a|G_0(f_{a'})_a$ 
for (a) $a'= (\pi {\rm N}_j{\rm N}_k){\rm N}_i$\ $a=(\pi {\rm N}_j)$,
and (b) $a'= (\pi {\rm N}_k)({\rm N}_i{\rm N}_j)$\ $a=({\rm N}_i{\rm N}_j)$.
The double full line describes the (NN) quasi--particle, while the
dashed--and--full double line is associated to the correlated ($\pi$N)
pair. The half--circles represent the two--body form factors.
\label{fig2}}
\end{figure}
\begin{figure}
\caption{The coupled equations (4.10) 
for Type--I partitions. The trapeziums represent the form factors
$|\tilde \Gamma_{a'a}>$ and $|\tilde \Gamma_{a'a_1}>$ for the 
virtual dissociation of the correlated $\pi$NN sub--system into a
particle and a correlated pair, or into two nucleons, respectively.
In the last graph, the presence of the one--pion--exchange tail in 
the NN potential ${\cal V}_{a_1}$ is shown. Only topologically 
different graphs are exhibited.
\label{fig3}}
\end{figure}
\begin{figure}
\caption{Graphical representation of the coupled equations (4.8)
for Type--II partitions. In this case the form factors 
$|\tilde {\bf \Gamma}^{(3)}_{a'}>$ (empty circles) describe
the virtual decay of a correlated pair in presence of the off--shell
propagation of the other pair, or of a virtual ($\pi$N)$\rightarrow$
N transition. The boxes enclose the terms which disappear
when the coupling between the $\pi$NNN and NNN spaces is switched
off in Type--II partitions.
\label{fig4}}
\end{figure}
\begin{figure}
\caption{The driving terms ${\sf Z}^{(2)}_{a'b'}$ for the quasi--particle 
Eqs. (4.14), as given by Eq. (4.15b). The driving terms obtainable from 
those given here by interchanging the initial and final states are not 
given. The dashed box encloses the contribution which has to be disregarded
to get connected--kernel equations. The full--line boxes exhibit
the terms which disappear when emission/absorption is ignored in 
$2+2$ partitions.
\label{fig5}}
\end{figure}
\begin{figure}
\caption{The non--vanishing contribution $<jk|G_0f(i)g_0f^\dagger(i)G_0|jk>$
to the operator $<a|G_0(f_{a'})_ag_0(f^\dagger_{a'})_bG_0|b>$. It
describes the virtual decay and subsequent formation of the NN cluster
$(jk)$, while the pion is absorbed and re--emitted by nucleon 
${\rm N}_i$.
\label{fig6}}
\end{figure}
\begin{figure}
\caption{The Eq. (4.18), expressing the $\pi + {\rm T}\rightarrow 
{\rm N} + {\rm N} + {\rm N }$ absorption amplitude in terms of the 
$2\rightarrow2$ transition amplitudes in the quasi-particle 
approximation.
\label{fig7}}
\end{figure}
\newpage
\begin{picture}(450,450)(20,0)
\thicklines
\multiput(70,390)(0,-40){3}{\line(5,0){300}}
\multiput(160,310)(120,0){2}{\circle*{5}}
\put(220,370){\circle{40}}
\multiput(165,305)(20,-20){2}{\line(1,-1){10}}
\multiput(245,275)(20,20){2}{\line(1,1){10}}
\multiput(205,270)(20,0){2}{\line(5,0){10}}
\put(206,230){{\LARGE (a)}}
\multiput(20,170)(0,-40){3}{\line(5,0){400}}
\multiput(70,90)(120,0){2}{\circle*{5}}
\multiput(250,90)(120,0){2}{\circle*{5}}
\multiput(130,150)(180,0){2}{\circle{40}}
\multiput(75,85)(20,-20){2}{\line(1,-1){10}}
\multiput(155,55)(20,20){2}{\line(1,1){10}}
\multiput(115,50)(20,0){2}{\line(5,0){10}}
\multiput(255,85)(20,-20){2}{\line(1,-1){10}}
\multiput(335,55)(20,20){2}{\line(1,1){10}}
\multiput(295,50)(20,0){2}{\line(5,0){10}}
\multiput(35,165)(180,0){2}{\multiput(0,0)(0,-20){2}{\oval(10,10)[l]}}
\multiput(35,155)(180,0){2}{\multiput(0,0)(0,-20){2}{\oval(10,10)[r]}}
\put(206,10){{\LARGE (b)}}
\end{picture}
\vfill
\centerline{{\bf Fig. 1}}
\newpage
\begin{picture}(450,450)(20,0)
\thicklines
\multiput(70,310)(0,40){2}{\line(5,0){300}}
\put(170,390){\line(5,0){200}}
\put(270,350){\circle*{5}}
\put(170,380){\oval(20,20)[l]}
\put(170,370){\line(0,5){20}}
\multiput(170,370)(20,-4){5}{\line(5,-1){10}}
\put(70,382){\line(5,0){90}}
\multiput(70,378)(20,0){5}{\line(5,0){10}}
\put(206,260){{\LARGE (a)}}
\put(70,130){\line(5,0){300}}
\multiput(170,170)(0,40){2}{\line(5,0){200}}
\put(270,130){\circle*{5}}
\put(170,190){\oval(40,40)[l]}
\put(170,170){\line(0,5){40}}
\multiput(70,188)(0,4){2}{\line(5,0){80}}
\multiput(70,90)(20,0){8}{\line(5,0){13}}
\multiput(227,95)(20,16){2}{\line(5,4){10}}
\put(206,40){{\LARGE (b)}}
\end{picture}
\vfill
\centerline{{\bf Fig. 2}}
\newpage
\begin{picture}(450,500)(20,0)
\thicklines
\put(0,400){\begin{picture}(450,100)(0,0)
\multiput(0,70)(0,-4){2}{\line(5,0){40}}
\multiput(0,30)(20,0){2}{\line(5,0){10}}
\multiput(40,30)(180,0){3}{\line(0,5){40}}
\multiput(60,45)(180,0){3}{\line(0,5){10}}
\multiput(40,30)(180,0){3}{\line(4,3){20}}
\multiput(40,70)(180,0){3}{\line(4,-3){20}}
\multiput(60,45)(180,0){3}{\line(5,0){20}}
\multiput(60,55)(180,0){3}{\line(5,0){20}}
\multiput(70,50)(180,0){3}{\line(5,0){10}}
\multiput(90,55)(0,-10){2}{\line(5,0){10}}
\multiput(110,62)(180,0){2}{\multiput(0,0)(0,-4){2}{\line(5,0){40}}}
\multiput(160,60)(180,0){2}{\oval(20,20)[l]}
\multiput(160,50)(180,0){2}{\line(0,5){20}}
\multiput(110,22)(20,0){4}{\line(5,0){10}}
\multiput(290,30)(20,0){4}{\line(5,0){10}}
\put(180,32){\oval(20,20)[r]}
\put(180,22){\line(0,5){20}}
\put(270,50){\line(5,0){10}}
\put(275,45){\line(0,5){10}}
\put(360,30){\circle*{5}}
\put(160,70){\line(5,0){60}}
\put(190,30){\line(5,0){30}}
\multiput(190,34)(20,0){2}{\line(5,0){10}}
\put(160,50){\line(5,-2){20}}
\put(340,70){\line(5,0){60}}
\put(360,30){\line(5,0){40}}
\put(340,50){\line(1,-1){20}}
\end{picture}}
\put(0,300){\begin{picture}(450,100)(0,0)
\put(0,70){\line(5,0){40}}
\put(0,30){\line(5,0){40}}
\multiput(0,34)(20,0){2}{\line(5,0){10}}
\multiput(40,30)(180,0){3}{\line(0,5){40}}
\multiput(60,45)(180,0){3}{\line(0,5){10}}
\multiput(40,30)(180,0){3}{\line(4,3){20}}
\multiput(40,70)(180,0){3}{\line(4,-3){20}}
\multiput(60,45)(180,0){3}{\line(5,0){20}}
\multiput(60,55)(180,0){3}{\line(5,0){20}}
\multiput(70,50)(180,0){3}{\line(5,0){10}}
\multiput(90,55)(0,-10){2}{\line(5,0){10}}
\put(110,78){\line(5,0){70}}
\put(110,38){\line(5,0){40}}
\multiput(110,42)(20,0){2}{\line(5,0){10}}
\multiput(180,68)(180,0){2}{\oval(20,20)[r]}
\multiput(180,58)(180,0){2}{\line(0,5){20}}
\multiput(160,40)(180,0){2}{\oval(20,20)[l]}
\multiput(160,30)(180,0){2}{\line(0,5){20}}
\put(270,50){\line(5,0){10}}
\put(275,45){\line(0,5){10}}
\multiput(190,70)(0,-4){2}{\line(5,0){30}}
\multiput(160,30)(20,0){3}{\line(5,0){10}}
\put(160,50){\line(5,2){20}}
\put(290,78){\line(5,0){70}}
\put(290,38){\line(5,0){40}}
\multiput(290,42)(20,0){2}{\line(5,0){10}}
\put(370,70){\line(5,0){30}}
\multiput(370,66)(20,0){2}{\line(5,0){10}}
\put(340,30){\line(5,0){60}}
\put(344,52){\line(5,2){10}}
\end{picture}}
\put(0,220){\begin{picture}(450,80)(0,0)
\thicklines
\put(220,30){\line(0,5){40}}
\put(240,45){\line(0,5){10}}
\put(220,30){\line(4,3){20}}
\put(220,70){\line(4,-3){20}}
\put(240,55){\line(5,0){20}}
\put(240,45){\line(5,0){20}}
\put(250,50){\line(5,0){10}}
\put(110,70){\line(5,0){110}}
\put(110,38){\line(5,0){40}}
\multiput(110,42)(20,0){2}{\line(5,0){10}}
\put(160,40){\oval(20,20)[l]}
\put(160,30){\line(0,5){20}}
\put(90,50){\line(5,0){10}}
\put(95,45){\line(0,5){10}}
\put(160,30){\line(5,0){60}}
\put(165,55){\line(1,1){10}}
\put(180,70){\circle*{5}}
\end{picture}}
\put(0,130){\begin{picture}(450,90)(0,0)
\multiput(0,70)(0,-40){2}{\line(5,0){40}}
\multiput(40,30)(180,0){3}{\line(0,5){40}}
\multiput(60,45)(180,0){3}{\line(0,5){10}}
\multiput(40,30)(180,0){3}{\line(4,3){20}}
\multiput(40,70)(180,0){3}{\line(4,-3){20}}
\multiput(60,45)(180,0){3}{\line(5,0){20}}
\multiput(60,55)(180,0){3}{\line(5,0){20}}
\multiput(70,50)(180,0){3}{\line(5,0){10}}
\multiput(90,55)(0,-10){2}{\line(5,0){10}}
\put(110,78){\line(5,0){70}}
\put(110,30){\line(5,0){42}}
\multiput(180,68)(180,0){2}{\oval(20,20)[r]}
\multiput(180,58)(180,0){2}{\line(0,5){20}}
\put(152,30){\circle*{5}}
\put(332,30){\circle*{5}}
\put(270,50){\line(5,0){10}}
\put(275,45){\line(0,5){10}}
\multiput(190,70)(0,-4){2}{\line(5,0){30}}
\multiput(159,30)(20,0){3}{\line(5,0){10}}
\put(152,30){\line(1,1){28}}
\put(290,78){\line(5,0){70}}
\put(290,30){\line(5,0){110}}
\put(370,70){\line(5,0){30}}
\multiput(370,66)(20,0){2}{\line(5,0){10}}
\put(334,32){\line(1,1){10}}
\put(350,48){\line(1,1){10}}
\end{picture}}
\put(0,50){\begin{picture}(450,80)(0,0)
\thicklines
\put(220,30){\line(0,5){40}}
\put(240,45){\line(0,5){10}}
\put(220,30){\line(4,3){20}}
\put(220,70){\line(4,-3){20}}
\put(240,55){\line(5,0){20}}
\put(240,45){\line(5,0){20}}
\put(250,50){\line(5,0){10}}
\multiput(110,70)(0,-40){2}{\line(5,0){110}}
\put(90,50){\line(5,0){10}}
\put(95,45){\line(0,5){10}}
\multiput(140,65)(20,-20){2}{\line(1,-1){10}}
\put(135,70){\circle*{5}}
\put(175,30){\circle*{5}}
\end{picture}}
\end{picture}
\vfill
\centerline{{\bf Fig. 3}}
\newpage
\begin{picture}(450,450)(20,0)
\thicklines
\put(0,300){\begin{picture}(450,150)(0,0)
\multiput(0,122)(20,0){2}{\line(5,0){10}}
\put(0,118){\line(5,0){40}}
\put(65,122){\line(5,0){10}}
\put(60,118){\line(5,0){20}}
\multiput(120,122)(20,0){2}{\line(5,0){10}}
\put(120,118){\line(5,0){40}}
\put(235,122){\line(5,0){10}}
\put(230,118){\line(5,0){20}}
\put(50,120){\circle{20}}
\put(170,120){\oval(20,20)[l]}
\put(170,110){\line(0,5){20}}
\put(220,120){\circle{20}}
\multiput(180,130)(20,0){2}{\line(5,0){10}}
\put(170,110){\line(5,0){50}}
\multiput(90,93.5)(0,-5){2}{\line(5,0){10}}
\multiput(0,70)(0,-20){2}{\line(5,0){50}}
\put(50,60){\circle{20}}
\multiput(60,62)(0,-4){2}{\line(5,0){20}}
\multiput(120,70)(0,-20){2}{\line(5,0){40}}
\put(160,60){\oval(20,20)[r]}
\put(160,50){\line(0,5){20}}
\multiput(170,62)(0,-4){2}{\line(5,0){40}}
\multiput(230,62)(0,-4){2}{\line(5,0){20}}
\put(220,60){\circle{20}}
\end{picture}}
\put(0,150){\begin{picture}(450,150)(0,0)
\multiput(0,130)(20,0){3}{\line(5,0){10}}
\put(0,110){\line(5,0){50}}
\multiput(120,130)(20,0){2}{\line(5,0){10}}
\put(120,110){\line(5,0){40}}
\put(65,122){\line(5,0){10}}
\put(60,118){\line(5,0){20}}
\multiput(170,122)(20,0){2}{\line(5,0){10}}
\put(170,118){\line(5,0){40}}
\put(235,122){\line(5,0){10}}
\put(230,118){\line(5,0){20}}
\put(50,120){\circle{20}}
\put(50,60){\circle{20}}
\put(160,120){\oval(20,20)[r]}
\put(160,110){\line(0,5){20}}
\put(220,120){\circle{20}}
\multiput(90,93.5)(0,-5){2}{\line(5,0){10}}
\put(170,60){\oval(20,20)[l]}
\put(170,50){\line(0,5){20}}
\multiput(170,70)(0,-20){2}{\line(5,0){50}}
\multiput(0,62)(0,-4){2}{\line(5,0){40}}
\multiput(60,62)(0,-4){2}{\line(5,0){20}}
\multiput(120,62)(0,-4){2}{\line(5,0){40}}
\multiput(230,62)(0,-4){2}{\line(5,0){20}}
\put(220,60){\circle{20}}
\put(260,90){\line(5,0){10}}
\put(265,85){\line(0,5){10}}
\multiput(280,120)(20,0){2}{\line(5,0){10}}
\put(280,100){\line(5,0){20}}
\put(300,100){\line(1,1){20}}
\put(320,120){\circle*{5}}
\put(320,120){\line(5,0){50}}
\put(380,120){\circle{20}}
\put(395,122){\line(5,0){10}}
\put(390,118){\line(5,0){20}}
\multiput(280,62)(0,-4){2}{\line(5,0){40}}
\multiput(390,62)(0,-4){2}{\line(5,0){20}}
\put(330,60){\oval(20,20)[l]}
\put(330,50){\line(0,5){20}}
\multiput(330,70)(0,-20){2}{\line(5,0){50}}
\put(380,60){\circle{20}}
\multiput(270,40)(0,100){2}{\line(5,0){150}}
\multiput(270,40)(150,0){2}{\line(0,5){100}}
\end{picture}}
\put(0,0){\begin{picture}(450,150)(0,0)
\put(0,120){\line(5,0){40}}
\multiput(0,70)(0,-20){2}{\line(5,0){50}}
\put(50,120){\circle{20}}
\put(50,60){\circle{20}}
\put(65,122){\line(5,0){10}}
\put(60,118){\line(5,0){20}}
\put(235,122){\line(5,0){10}}
\put(230,118){\line(5,0){20}}
\multiput(90,93.5)(0,-5){2}{\line(5,0){10}}
\put(120,130){\line(5,0){40}}
\put(160,130){\circle*{5}}
\multiput(170,130)(20,0){3}{\line(5,0){10}}
\put(160,130){\line(1,-1){20}}
\put(180,110){\line(5,0){40}}
\put(220,120){\circle{20}}
\multiput(120,70)(0,-20){2}{\line(5,0){40}}
\put(160,60){\oval(20,20)[r]}
\put(160,50){\line(0,5){20}}
\multiput(60,62)(0,-4){2}{\line(5,0){20}}
\multiput(170,62)(0,-4){2}{\line(5,0){40}}
\multiput(230,62)(0,-4){2}{\line(5,0){20}}
\put(220,60){\circle{20}}
\multiput(110,40)(0,100){2}{\line(5,0){150}}
\multiput(110,40)(150,0){2}{\line(0,5){100}}
\end{picture}}
\end{picture}
\vfill
\centerline{{\bf Fig. 4}}
\newpage
\begin{picture}(450,550)(40,0)
\thicklines
\put(0,0){\begin{picture}(450,70)(0,0)
\thicklines
\multiput(95,70)(10,0){3}{\line(5,0){5}}
\put(95,66){\line(5,0){30}}
\multiput(95,30)(0,4){2}{\line(5,0){30}}
\multiput(165,70)(0,-4){2}{\line(5,0){30}}
\multiput(170,34)(10,0){3}{\line(5,0){5}}
\put(165,30){\line(5,0){30}}
\put(125,70){\line(5,0){40}}
\put(125,30){\line(0,5){40}}
\put(125,30){\line(5,0){40}}
\put(165,30){\line(0,5){40}}
\multiput(205,55)(0,-10){2}{\line(5,0){10}}
\multiput(225,72)(10,0){3}{\line(5,0){5}}
\put(225,68){\line(5,0){30}}
\multiput(225,38)(0,4){2}{\line(5,0){30}}
\multiput(265,70)(0,-30){2}{\circle{20}}
\multiput(315,60)(0,-30){2}{\circle{20}}
\put(275,70){\line(5,0){40}}
\put(265,50){\line(5,0){50}}
\put(265,30){\line(5,0){40}}
\multiput(325,62)(0,-4){2}{\line(5,0){30}}
\multiput(330,32)(10,0){3}{\line(5,0){5}}
\put(325,28){\line(5,0){30}}
\multiput(85,14)(0,70){2}{\line(5,0){280}}
\multiput(85,14)(280,0){2}{\line(0,5){70}}
\put(218,-5){{\Large (f)}}
\end{picture}}
\put(0,90){\begin{picture}(450,70)(0,0)
\thicklines
\put(20,70){\line(5,0){70}}
\multiput(20,30)(0,10){2}{\line(5,0){30}}
\multiput(20,35)(10,0){3}{\line(5,0){5}}
\multiput(90,70)(0,-4){2}{\line(5,0){30}}
\multiput(95,34)(10,0){3}{\line(5,0){5}}
\put(90,30){\line(5,0){30}}
\put(50,30){\line(0,5){40}}
\put(90,30){\line(0,5){40}}
\put(50,30){\line(5,0){40}}
\multiput(130,55)(0,-10){2}{\line(5,0){10}}
\multiput(150,75)(155,0){2}{\line(5,0){85}}
\multiput(150,50)(155,0){2}{\line(5,0){30}}
\multiput(150,45)(155,0){2}{\multiput(0,0)(10,0){3}{\line(5,0){5}}}
\multiput(150,40)(155,0){2}{\line(5,0){30}}
\multiput(180,40)(155,0){2}{\line(0,5){10}}
\multiput(195,35)(155,0){2}{\line(0,5){20}}
\multiput(180,50)(155,0){2}{\line(3,1){15}}
\multiput(180,40)(155,0){2}{\line(3,-1){15}}
\multiput(235,65)(0,-28){2}{{\circle{20}}}
\multiput(390,65)(0,-30){2}{{\circle{20}}}
\multiput(195,55)(155,0){2}{\line(5,0){40}}
\multiput(195,39)(10,0){3}{\line(5,0){5}}
\put(195,35){\line(5,0){30}}
\put(350,35){\line(5,0){30}}
\multiput(245,67)(155,0){2}{\multiput(0,0)(0,-4){2}{\line(5,0){30}}}
\multiput(250,39)(10,0){3}{\line(5,0){5}}
\put(245,35){\line(5,0){30}}
\multiput(405,37)(10,0){3}{\line(5,0){5}}
\put(400,33){\line(5,0){30}}
\put(285,50){\line(5,0){10}}
\put(290,45){\line(0,5){10}}
\multiput(295,20)(0,160){2}{\line(5,0){145}}
\multiput(295,20)(145,0){2}{\line(0,5){160}}
\multiput(312,104)(0,73){2}{\multiput(0,0)(20,0){6}{\line(5,0){10}}}
\multiput(300,105)(135,0){2}{\multiput(0,0)(0,20){4}{\line(0,5){10}}}
\put(218,7){{\Large (e)}}
\end{picture}}
\put(0,180){\begin{picture}(450,70)(0,0)
\thicklines
\multiput(20,70)(0,-10){2}{\line(5,0){30}}
\multiput(20,65)(10,0){3}{\line(5,0){5}}
\put(20,30){\line(5,0){70}}
\multiput(90,70)(0,-4){2}{\line(5,0){30}}
\multiput(95,34)(10,0){3}{\line(5,0){5}}
\put(90,30){\line(5,0){30}}
\put(50,30){\line(0,5){40}}
\put(90,30){\line(0,5){40}}
\put(50,70){\line(5,0){40}}
\multiput(130,55)(0,-10){2}{\line(5,0){10}}
\multiput(150,65)(0,-10){2}{\line(5,0){30}}
\multiput(150,60)(10,0){3}{\line(5,0){5}}
\put(150,30){\line(5,0){85}}
\put(180,55){\line(0,5){10}}
\put(195,50){\line(0,5){20}}
\put(180,65){\line(3,1){15}}
\put(180,55){\line(3,-1){15}}
\multiput(235,68)(0,-28){2}{\circle{20}}
\multiput(195,70)(0,-4){2}{\line(5,0){30}}
\multiput(200,50)(10,0){3}{\line(5,0){5}}
\multiput(245,70)(0,-4){2}{\line(5,0){30}}
\put(245,38){\line(5,0){30}}
\multiput(250,42)(10,0){3}{\line(5,0){5}}
\put(285,50){\line(5,0){10}}
\put(290,45){\line(0,5){10}}
\multiput(305,73)(0,-10){2}{\line(5,0){30}}
\multiput(305,68)(10,0){3}{\line(5,0){5}}
\put(305,30){\line(5,0){75}}
\put(335,63){\line(0,5){10}}
\put(350,58){\line(0,5){20}}
\put(335,73){\line(3,1){15}}
\put(335,63){\line(3,-1){15}}
\put(390,68){\circle{20}}
\put(390,30){\circle{20}}
\multiput(350,78)(0,-20){2}{\line(5,0){40}}
\multiput(400,70)(0,-4){2}{\line(5,0){30}}
\multiput(405,32)(10,0){3}{\line(5,0){5}}
\put(400,28){\line(5,0){30}}
\put(218,6){{\Large (d)}}
\end{picture}}
\put(0,270){\begin{picture}(450,70)(0,0)
\thicklines
\multiput(97.5,70)(0,-5){3}{\line(5,0){30}}
\multiput(97.5,30)(10,0){3}{\line(5,0){5}}
\multiput(167.5,70)(0,-4){2}{\line(5,0){30}}
\put(167.5,30){\line(5,0){30}}
\multiput(172.5,34)(10,0){3}{\line(5,0){5}}
\put(127.5,30){\line(0,5){40}}
\put(167.5,30){\line(0,5){40}}
\put(127.5,70){\line(5,0){40}}
\put(127.5,30){\line(5,0){40}}
\multiput(207.5,55)(0,-10){2}{\line(5,0){10}}
\multiput(227.5,65)(0,-5){3}{\line(5,0){30}}
\multiput(227.5,30)(10,0){8}{\line(5,0){5}}
\put(257.5,55){\line(0,5){10}}
\put(272.5,50){\line(0,5){20}}
\put(257.5,65){\line(3,1){15}}
\put(257.5,55){\line(3,-1){15}}
\multiput(312.5,68)(0,-28){2}{\circle{20}}
\multiput(272.5,70)(0,-4){2}{\line(5,0){30}}
\put(272.5,50){\line(5,0){40}}
\multiput(322.5,70)(0,-4){2}{\line(5,0){30}}
\put(322.5,38){\line(5,0){30}}
\multiput(327.5,42)(10,0){3}{\line(5,0){5}}
\put(218,6){{\Large (c)}}
\end{picture}}
\put(0,360){\begin{picture}(450,70)(0,0)
\thicklines
\put(15,70){\line(5,0){100}}
\put(15,40){\line(5,0){30}}
\multiput(15,35)(10,0){3}{\line(5,0){5}}
\put(15,30){\line(5,0){100}}
\multiput(85,65)(10,0){3}{\line(5,0){5}}
\put(85,60){\line(5,0){30}}
\put(45,30){\line(0,5){40}}
\put(85,30){\line(0,5){40}}
\multiput(125,45)(0,10){2}{\line(5,0){10}}
\put(285,50){\line(5,0){10}}
\put(290,45){\line(0,5){10}}
\multiput(145,70)(160,0){2}{\line(5,0){85}}
\put(145,49){\line(5,0){30}}
\put(305,45){\line(5,0){30}}
\put(145,44){\multiput(0,0)(10,0){3}{\line(5,0){5}}}
\put(305,40){\multiput(0,0)(10,0){3}{\line(5,0){5}}}
\put(145,39){\line(5,0){30}}
\put(305,35){\line(5,0){30}}
\put(175,39){\line(0,5){10}}
\put(190,34){\line(0,5){20}}
\put(175,49){\line(3,1){15}}
\put(175,39){\line(3,-1){15}}
\put(335,35){\line(0,5){10}}
\put(350,30){\line(0,5){20}}
\put(335,45){\line(3,1){15}}
\put(335,35){\line(3,-1){15}}
\multiput(230,50)(160,0){2}{\line(0,5){20}}
\multiput(245,55)(160,0){2}{\line(0,5){10}}
\multiput(230,50)(160,0){2}{\line(3,1){15}}
\multiput(230,70)(160,0){2}{\line(3,-1){15}}
\put(190,34){\line(5,0){85}}
\put(350,30){\line(5,0){85}}
\multiput(245,65)(160,0){2}{\line(5,0){30}}
\multiput(245,60)(160,0){2}{\multiput(0,0)(10,0){3}{\line(5,0){5}}}
\multiput(245,55)(160,0){2}{\line(5,0){30}}
\put(350,50){\line(5,0){40}}
\put(190,50){\line(5,0){40}}
\multiput(190,54)(10,0){4}{\line(5,0){5}}
\put(218,6){{\Large (b)}}
\end{picture}}
\put(0,450){\begin{picture}(450,70)(0,0)
\thicklines
\multiput(97.5,70)(10,0){3}{\line(5,0){5}}
\put(97.5,40){\line(5,0){30}}
\put(97.5,35){\line(5,0){30}}
\put(97.5,30){\line(5,0){100}}
\put(127.5,70){\line(5,0){70}}
\multiput(167.5,65)(10,0){3}{\line(5,0){5}}
\put(167.5,60){\line(5,0){30}}
\put(127.5,30){\line(0,5){40}}
\put(167.5,30){\line(0,5){40}}
\multiput(207.5,45)(0,10){2}{\line(5,0){10}}
\multiput(227.5,70)(10,0){8}{\line(5,0){5}}
\put(227.5,49){\line(5,0){30}}
\put(227.5,44){\line(5,0){30}}
\put(227.5,39){\line(5,0){30}}
\put(257.5,39){\line(0,5){10}}
\put(272.5,34){\line(0,5){20}}
\put(257.5,49){\line(3,1){15}}
\put(257.5,39){\line(3,-1){15}}
\put(307.5,50){\line(0,5){20}}
\put(322.5,55){\line(0,5){10}}
\put(307.5,70){\line(3,-1){15}}
\put(307.5,50){\line(3,1){15}}
\multiput(272.5,54)(0,-4){2}{\line(5,0){35}}
\multiput(322.5,65)(0,-10){2}{\line(5,0){30}}
\multiput(322.5,60)(10,0){3}{\line(5,0){5}}
\put(272.5,34){\line(5,0){80}}
\put(218,5){{\Large (a)}}
\end{picture}}
\end{picture}
\vfill
\centerline{{\bf Fig. 5}}
\newpage
\begin{picture}(450,450)(20,0)
\thicklines
\multiput(70,302)(230,0){2}{\multiput(0,0)(0,-4){2}{\line(5,0){70}}}
\put(160,300){\oval(40,40)[l]}
\put(280,300){\oval(40,40)[r]}
\multiput(160,280)(120,0){2}{\line(0,5){40}}
\multiput(160,320)(0,-40){2}{\line(5,0){120}}
\put(70,230){\line(5,0){300}}
\multiput(180,230)(80,0){2}{\circle*{5}}
\multiput(125,175)(20,20){3}{\line(1,1){10}}
\multiput(265,225)(20,-20){3}{\line(1,-1){10}}
\end{picture}
\vfill
\centerline{{\bf Fig. 6}}
\newpage
\begin{picture}(450,450)(20,0)
\thicklines
\put(80,350){\line(5,0){40}}
\put(80,330){\line(5,0){20}}
\put(80,310){\line(5,0){80}}
\put(120,330){\circle{40}}
\put(135,315){\line(5,0){25}}
\put(140,320){\line(5,0){20}}
\multiput(160,350)(-10,0){4}{\line(-5,0){5}}
\multiput(180,335)(0,-10){2}{\line(5,0){10}}
\put(180,205){\line(5,0){10}}
\put(185,200){\line(0,5){10}}
\multiput(340,330)(0,-120){2}{\circle{40}}
\multiput(380,350)(0,-120){2}{\multiput(0,0)(-10,0){4}{\line(-5,0){5}}}
\multiput(380,320)(0,-120){2}{\line(-5,0){22}}
\multiput(380,315)(0,-120){2}{\line(-5,0){28}}
\multiput(380,310)(0,-120){2}{\line(-5,0){40}}
\put(270,310){\line(5,0){70}}
\multiput(270,315)(10,0){6}{\line(5,0){5}}
\put(270,320){\line(5,0){52}}
\thinlines
\multiput(273,309)(3,0){16}{\line(0,5){12}}
\put(321,309){\line(0,5){11}}
\put(324,309){\line(0,5){7}}
\put(327,309){\line(0,5){5}}
\put(330,309){\line(0,5){3}}
\thicklines
\put(270,190){\line(5,0){70}}
\put(270,194){\line(5,0){59}}
\put(270,230){\line(5,0){70}}
\multiput(270,226)(10,0){6}{\line(5,0){5}}
\thinlines
\multiput(273,231)(3,0){17}{\line(0,-5){6}}
\put(324,231){\line(0,-5){6}}
\put(327,231){\line(0,-5){4}}
\put(330,231){\line(0,-5){3}}
\thicklines
\put(250,295){\line(0,5){40}}
\put(270,310){\line(0,5){10}}
\put(250,295){\line(4,3){20}}
\put(250,335){\line(4,-3){20}}
\put(260,228){\circle{20}}
\put(260,192){\circle{20}}
\put(210,350){\line(5,0){130}}
\multiput(210,335)(0,-40){2}{\line(5,0){40}}
\put(210,228){\line(5,0){40}}
\multiput(210,202)(0,-20){2}{\line(5,0){50}}
\end{picture}
\vfill
\centerline{{\bf Fig. 7}}

\begin{references}
\bibitem{note1} To be rigorous, one ought to introduce integrations
over the momenta of the spectator particles in eq.~(\ref{pole}).
Since these integrations are removed when considering the 
momentum--space matrix elements of the relevant transition operators, to 
save on notation here and in the following we shall live such integrations 
implicit.
\bibitem{note2} Strictly speaking, because of the identity of the
nucleons, one has a unique transition amplitude for the process
$\pi + {\rm T} \rightarrow {\rm N} + {\rm D}$. As usual in scattering
theory, for the moment we assume the nucleons as distinguishable, so 
that the configurations $(\pi{\rm N}_j{\rm N}_k)$
and $(\pi{\rm N}_i{\rm N}_k)$ are regarded as
different from each other. The prescriptions due to particle identity can
be taken into account at the end, through suitable linear combinations
of the transition amplitudes. However, to save on notation, we shall
simply write ${\rm D}$ in place of 
$({\rm N}_i {\rm N}_k)$ in the transition amplitudes.
\end{references}
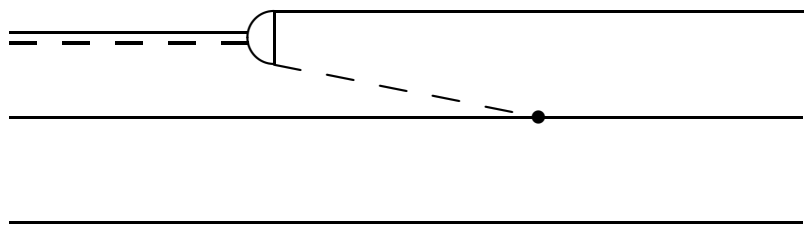
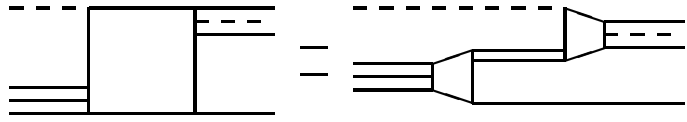
\end{document}